\documentclass[varenna]{cimento}
\usepackage{graphicx}
\def\gtsima{$\; \buildrel > \over \sim \;$}
\def\ltsima{$\; \buildrel < \over \sim \;$}
\def\gtrsim{\lower.5ex\hbox{\gtsima}}
\def\lesssim{\lower.5ex\hbox{\ltsima}}
\def\apjl{\apj Letter}
\def\nat{Nature}
\def\mnras{MNRAS}
\def\prd{Physical Review Documents}
\def\aap{Astronomy \& Astrophysics}
\def\pasa{Publications of the Astronomical Society of Australia}
\def\apjs{\apj Supplement}
\def\nar{New Astronomy}
\def\aapr{The Astronomy and Astrophysics Review}
\def\araa{Annual Review of Astronomy and Astrophysics}
\def\aj{Astronomical Journal}
\def\physrep{Physics Reports}
\def\planss{Planetary and Space Science}
\def\pasj{Publications of the Astronomical Society of Japan}
\title{Astrophysics of stellar black holes}
\author{Michela Mapelli$^{1,2,3}$}
\institute{$1$ Institut f\"ur Astro- und Teilchenphysik, Universit\"at Innsbruck, Technikerstrasse 25/8, A--6020 Innsbruck, Austria\\
  $2$ INAF -- Osservatorio Astronomico di Padova, Vicolo dell'Osservatorio 5, I--35142 Padova, Italy\\
$3$ INFN -- Sezione Padova, Via F.~Marzolo 8, I--35131 Padova, Italy}

\PACSes{\PACSit{--.--}{\dots}
\PACSit{--.--}{\ldots}}

\def\gtsima{$\; \buildrel > \over \sim \;$}
\def\ltsima{$\; \buildrel < \over \sim \;$}
\def\gtrsim{\lower.5ex\hbox{\gtsima}}
\def\lesssim{\lower.5ex\hbox{\ltsima}}
\def\apj{Astrophysical Journal}

\begin{document}

\maketitle

\begin{abstract}
On September 14 2015, the LIGO interferometers captured a gravitational wave (GW) signal from two merging black holes (BHs), opening the era of GW astrophysics. Five BH mergers have been reported so far, three  of them involving massive BHs ($\gtrsim{}30$ M$_\odot$). According to stellar evolution models, such massive BHs can originate from massive, relatively metal-poor stars. The formation channels of merging BH binaries are still an open question: a plethora of uncertainties affect the evolution of massive stellar binaries (e.g. the process of common envelope) and their dynamics. This review aims to discuss the open questions about BH binaries, and to present the state-of-the-art knowledge about the astrophysics of BHs to non-specialists, in light of the first LIGO-Virgo detections.
\end{abstract}

\section{Lesson learned from the first direct gravitational wave detections}
On September 14 2015, the LIGO interferometers captured a gravitational wave (GW) signal from two merging black holes (BHs, \cite{abbott2016a}). This event, named GW150914, is the first direct detection of GWs, about hundred years after Einstein's prediction \cite{einstein1916,einstein1918}. To date, four more BH mergers have been reported (GW151226, GW170104, GW170608 and GW170814 \cite{abbott2016b,abbott2017a,abbott2017c,abbott2017b}), plus a sixth candidate (LVT151012, \cite{abbott2016c}). In particular, GW170814 was detected jointly by three interferometers: the two LIGO detectors in the United States \cite{LIGOdetector} and Virgo in Italy \cite{Virgodetector}. Table~\ref{tab:table1} summarizes the main properties of the five published BH mergers and of the sixth candidate event. Moreover, during the second (and currently the last) observing run, the LIGO and Virgo interferometers have detected the first double neutron star (NS) merger: GW170817 \cite{abbott2017GW170817,abbott2017multimessenger}. 
\vspace{0.5cm}

Astrophysicists have learned several revolutionary concepts about BHs from GW detections \cite{abbott2016d}. First, GW150914 has confirmed the existence of double BH binaries (BHBs), i.e. binaries composed of two BHs. BHBs have been predicted a long time ago (e.g. \cite{tutukov1973,thorne1987,schutz1989,kulkarni1993,sigurdsson1993,bethe1998,portegieszwart2000,colpi2003,belczynski2004}), but their observational confirmation was still missing. Second, GW detections show that some BHBs are able to merge within a Hubble time.

\begin{figure}
\center{
\includegraphics[width=12cm]{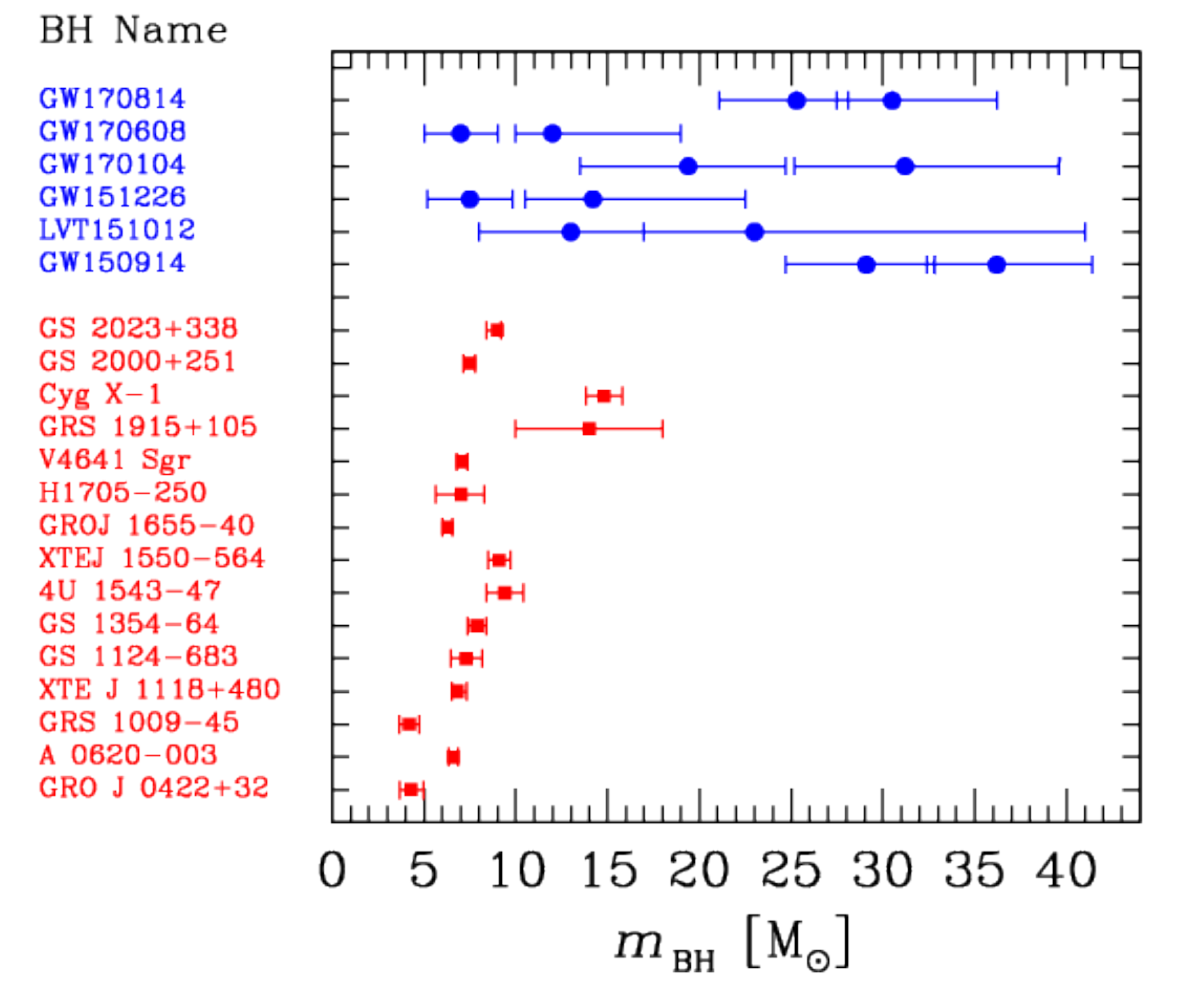}     % includes figure foo.eps
}
\caption{\label{fig:fig1}A compilation of BH masses $m_{\rm BH}$ from observations. Red squares: BHs with dynamical mass measurement in X-ray binaries \cite{orosz2003,ozel2010}. This selected sample is quite conservative, because uncertain and debated results are not being shown (e.g. IC10 X-1 \cite{laycock2015}). Blue circles: BHs in the first published GW events \cite{abbott2016c,abbott2017a,abbott2017b}.}
\end{figure}

Finally, three out of five merging BHBs detected so far (GW150914, GW170104 and GW170814) host BHs with mass in excess of 20 M$_\odot$. This result was a genuine surprise for the astrophysicists, because the only stellar BHs for which we have a dynamical mass measurement, i.e. about a dozen of BHs in X-ray binaries, have  mass  $<20$~M$_{\odot}$ (see Figure~\ref{fig:fig1} for a compilation of measured BH masses). Moreover, most theoretical models did not predict the existence of BHs with mass $m_{\rm BH}>30$ M$_\odot$ (but see \cite{mapelli2009,mapelli2010,belczynski2010,fryer2012,mapelli2013,ziosi2014,spera2015} for several exceptions). % (Figure~\ref{fig:fig2}). 
Thus, the first GW detections have urged the astrophysical community to deeply revise the models of BH formation and evolution.

%%%%%%%%%%%%%%%%%%%%%%%%%%%%%FIGURE 1%%%%%%%%%%%%%%%%%%%%%%%%%%%%%%%%%%%%%%%%
%%%%%%%%%%%%%%%%%%%%%%%%%%%%%%%%%%%%%%%%%%%%%%%%%%%%%%%%%%%%%%%%%%%%%%%%%%%%

%%%%%%%%%%%%%%%%%%%%%%%%%%%%%FIGURE 2%%%%%%%%%%%%%%%%%%%%%%%%%%%%%%%%%%%%%%%%
%\begin{figure}
%\center{
%\includegraphics[width=12cm]{BSE}     % includes figure foo.eps
%}
%\caption{\label{fig:fig2}Mass of the compact remnant ($M_{\rm rem}$) as a function of the zero-age main sequence (ZAMS) mass (i.e. the initial mass of the progenitor star) for four different metallicities ($Z=0.0002,$ 0.002, 0.01 and 0.02). These BH mass functions are obtained with the {\sc SSE} code, which is one of the most popular open-source population-synthesis codes  \cite{hurley2000}. The red and the blue horizontal strips show the mass of primary and secondary BHs in GW150914.}
%\end{figure}
%%%%%%%%%%%%%%%%%%%%%%%%%%%%%%%%%%%%%%%%%%%%%%%%%%%%%%%%%%%%%%%%%%%%%%%%%%%%

This review discusses the formation channels of BHs and BHBs in light of the challenges posed by recent GW detections. It is aimed at students and non-expert in the field, being the proceeding of the lecture held for the Course 200 on ``Gravitational Waves and Cosmology'' of the International School of Physics ``Enrico Fermi''.

%%%%%%%%%%%%%%%%%TABLE%%%%%%%%%%%%
\begin{table}[h]
   \caption{Main observed properties of published GW detections (BH mergers only).  \label{tab:table1}}
  \begin{center}
    \begin{tabular}{lllllll}
      \hline
      & GW150914             & LVT151012           & GW151226                & GW170104               &  GW170608       & GW170814\\
      \hline
%\mr
$m_1$ (M$_\odot$)         & $36.2^{+5.2}_{-3.8}$    & $23^{+18}_{-6}$       & $14.2^{+8.3}_{-3.7}$   &   $31.2^{+8.4}_{-6.0}$  & $12^{+7}_{-2}$   & $30.5^{+5.7}_{-3.0}$ \vspace{0.1cm}\\
$m_2$ (M$_\odot$)         & $29.1^{+3.7}_{-4.4}$    & $13^{+4}_{-5}$        & $7.5^{+2.3}_{-2.3}$    &    $19.4^{+5.3}_{-5.9}$ & $7^{+2}_{-2}$    & $25.3^{+2.8}_{-4.2}$ \vspace{0.1cm}\\
$m_{\rm chirp}$ (M$_\odot$)  & $28.1^{+1.8}_{-1.5}$    & $15.1^{+1.4}_{-1.1}$  & $8.9^{+0.3}_{-0.3}$    &   $21.1^{+2.4}_{-2.7}$  & $7.9^{+0.2}_{-0.2}$ & $24.1^{+1.4}_{-1.1}$ \vspace{0.1cm}\\
$m_{\rm TOT}$ (M$_\odot$)           & $65.3^{+4.1}_{-3.4}$    & $37^{+13}_{-4}$       & $21.8^{+5.9}_{-1.7}$   &    $50.7^{+5.9}_{-5.0}$ & $19^{+5}_{-1}$ & $55.9^{+3.4}_{-2.7}$ \vspace{0.1cm}\\
$m_{\rm fin}$ (M$_\odot$)   & $62.3^{+3.7}_{-3.1}$    & $35^{+14}_{-4}$       & $20.8^{+6.1}_{-1.7}$   &    $48.7^{+5.7}_{-4.6}$ & $18.0^{+4.8}_{-0.9}$ & $53.2^{+3.2}_{-2.5}$ \vspace{0.3cm}\\
$\chi{}_{\rm eff}$         & $-0.06^{+0.14}_{-0.14}$ & $0.0^{+0.3}_{-0.2}$    & $0.21^{+0.20}_{-0.10}$ &  $-0.12^{+0.21}_{-0.30}$ & $0.07^{+0.23}_{-0.09}$ & $0.06^{+0.12}_{-0.12}$ \vspace{0.1cm}\\
$S_{\rm f}$               & $0.68^{+0.05}_{-0.06}$   & $0.66^{+0.09}_{-0.10}$ & $0.74^{+0.06}_{-0.06}$ &  $0.64^{+0.09}_{-0.20}$  & $0.69^{+0.04}_{-0.05}$ & $0.70^{+0.07}_{-0.05}$ \vspace{0.3cm}\\
$d_{\rm L}$ (Mpc)         & $420^{+150}_{-180}$     & $1000^{+500}_{-500}$   & $440^{+180}_{-190}$    &  $880^{+450}_{-390}$    & $340^{+140}_{-140}$ & $540^{+130}_{-210}$ \vspace{0.1cm}\\
$z$                     & $0.09^{+0.03}_{-0.04}$   & $0.20^{+0.09}_{-0.09}$ & $0.09^{+0.03}_{-0.04}$  &   $0.18^{+0.08}_{-0.07}$ & $0.07^{+0.03}_{-0.03}$ & $0.11^{+0.03}_{-0.04}$\vspace{0.1cm}\\
\hline
%\br
    \end{tabular}
  \end{center}
  \footnotesize{$m_1$: mass of the primary BH; $m_2$: mass of the secondary BH; $m_{\rm chirp}=m_1^{3/5}\,{}m_2^{3/5}\,{}(m_1+m_2)^{-1/5}$: chirp mass of the binary; $m_{\rm tot}=m_1+m_2$: total mass of the binary; $m_{\rm fin}$: mass of the final BH; $\chi_{\rm eff}$: effective spin (defined in equation~\ref{eq:effectivespin}); $S_{\rm fin}$: spin of the final BH; $d_{\rm L}$: luminosity distance; $z$: redshift. From left to right: GW150914, LVT151012, GW151226, GW170608, GW170104, GW170814. For all properties, we report median values with 90\%{} credible intervals (\cite{abbott2016c,abbott2017a,abbott2017b,abbott2017c}). Source-frame masses are quoted. }
\end{table}
%%%%%%%%%%%%%%%%%%%%%%%%

\section{The formation of compact remnants from stellar evolution and supernova explosions}
BHs and NSs are expected to form as remnants of massive ($\gtrsim{}8$ M$_\odot$) stars. An alternative theory predicts that BHs can form also from gravitational collapse in the early Universe (the so called primordial BHs, e.g. \cite{bird2016,carr2016,inomata2016}). In this review, we will focus on BHs of stellar origin.

The mass function of BHs is highly uncertain, because it may be affected by a number of barely understood processes. In particular, stellar winds and supernova (SN) explosions both play a major role on the formation of compact remnants. Processes occurring in close binary systems (e.g. mass transfer and common envelope) are a further complication and will be discussed in the next section.  

\subsection{Stellar winds and stellar evolution}
Stellar winds are outflows of gas from the atmosphere of a star. In cold stars (e.g. red giants and asymptotic giant branch stars) they are mainly induced by radiation pressure on dust, which forms in the cold outer layers (e.g. \cite{vanloon2005}). In massive hot stars (O and B main sequence stars, luminous blue variables and Wolf-Rayet stars), stellar winds are powered by the coupling between the momentum of photons and that of metal ions present in the stellar photosphere. A large number of strong and weak resonant metal lines are responsible for this coupling (see e.g. \cite{bresolin2004} for a review).

Understanding stellar winds is tremendously important for the study of compact objects, because mass loss determines the pre-SN mass of a star (both its total mass and its core mass), which in turn affects the outcome of an SN explosion \cite{fryer1999,fryer2001,mapelli2009,belczynski2010}. 

% The theory of stellar winds and our knowledge of massive star evolution have been deeply revised in the last decades. 
Early work on stellar winds (e.g. \cite{abbott1982,kudritzki1987,leitherer1992}) highlighted that the mass loss of O and B stars depends on metallicity as $\dot{m}\propto{}Z^\alpha$ (with $\alpha{}\sim{}0.5-1.0$, depending on the model). However, such early work did not account for multiple scattering, i.e. for the possibility that a photon interacts several times before being absorbed or leaving the photosphere. Vink et al. (2001, \cite{vink2001}) accounted for multiple scatterings and found a  universal metallicity dependence $\dot{m}\propto{}Z^{0.85}\,{}v_\infty^p$, where $v_\infty$ is the terminal velocity\footnote{The terminal velocity $v_\infty$ is the the velocity reached by the wind at large distance from the star, where the radiative acceleration approaches zero because of the geometrical dilution of the photospheric radiation field. Since line-driven winds are continuously accelerated by the absorption of photospheric photons in spectral lines, $v_\infty$ corresponds to the maximum velocity of the stellar wind. See the review by Kudritzki \& Puls (2000, \cite{kudritzki2000}) for more details.}  and $p=-1.23$ ($p=-1.60$) for stars with effective temperature $T_{\rm eff}\gtrsim{}25000$ K ($12000\,{}{\rm K}\lesssim{}T_{\rm eff}\lesssim{}25000$~K).

The situation is more uncertain for post-main sequence stars. For Wolf-Rayet (WR) stars, i.e. naked Helium cores, \cite{vinkdekoter2005} predict a similar trend with metallicity $\dot{m}\propto{}Z^{0.86}$. With a different numerical approach (which accounts also for wind clumping), \cite{graefener2008} find a strong dependence of WR mass loss on metallicity but also on the electron-scattering Eddington factor $\Gamma_e=\kappa_e\,{}L\,{}/(4\,{}\pi{}\,{}c\,{}G\,{}m)$, where $\kappa_e$ is the cross section for electron scattering, $L$ is the stellar luminosity, $c$ is the speed of light, $G$ is the gravity constant, and $m$ is the stellar mass. The importance of $\Gamma_e$ has become increasingly clear in the last few years \cite{graefener2011,vink2011,vink2016}, but, unfortunately, only few stellar evolution models include this effect.

\begin{figure}
\center{
\includegraphics[width=12cm]{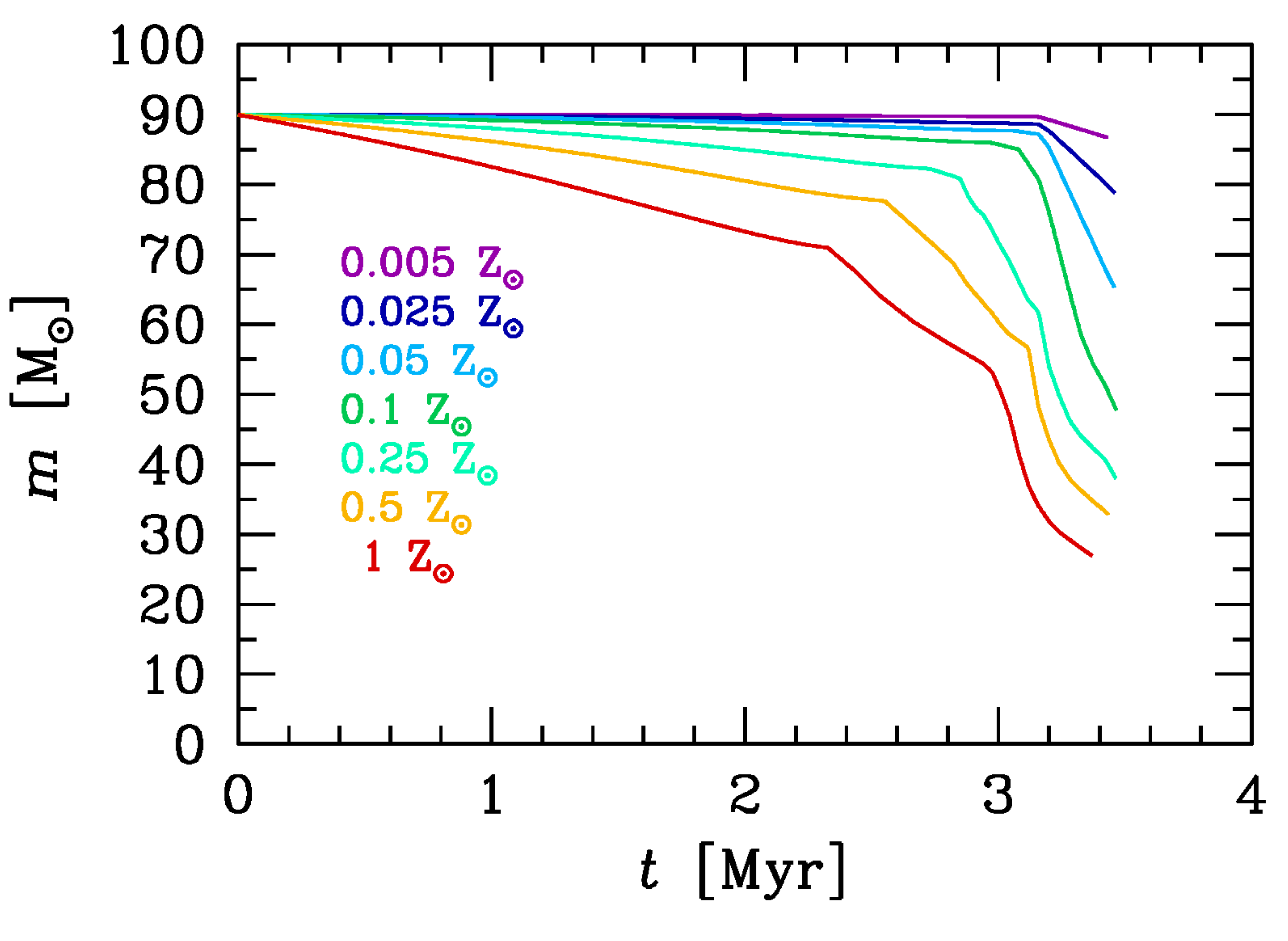}     % includes figure foo.eps
}
\caption{\label{fig:massloss}Evolution of stellar mass as a function of time for a star with ZAMS mass $m_{\rm ZAMS}=90$~M$_\odot$ and seven different metallicities, ranging from 0.005~Z$_\odot{}$ up to Z$_\odot$ (we assumed $Z_\odot=0.02$). These curves were obtained with the {\sc SEVN} population-synthesis code \cite{spera2015}, adopting {\sc PARSEC} stellar evolution tracks \cite{chen2015}.}
\end{figure}

For example, \cite{tang2014,chen2015} adopt a mass loss prescriptions $\dot{m}\propto{}Z^\alpha$, where $\alpha{}=0.85$ if $\Gamma_e<2/3$ and $\alpha{}=2.45-2.4\,{}\Gamma_e$ if $2/3\leq{}\Gamma_e\leq{}1$. This simple formula accounts for the fact that metallicity dependence tends to vanish when the star is close to be radiation pressure dominated, as clearly shown by figure 10 of \cite{graefener2008}. Figure~\ref{fig:massloss} shows the mass evolution of a star with zero-age main sequence (ZAMS) mass $m_{\rm ZAMS}=90$ M$_\odot$ for seven different metallicities, as obtained with the {\sc SEVN} code \cite{spera2015}. At the end of its life, a solar-metallicity star (here we assume $Z_\odot=0.02$) has lost more than $2/3$ of its initial mass, while the most metal-poor star in the Figure ($Z=0.005$ Z$_\odot$) has retained almost all its initial mass.

%%%%%%%%%%%%%%%%%%%%%%%%%%%%%FIGURE %%%%%%%%%%%%%%%%%%%%%%%%%%%%%%%%%%%%%%%%
%%%%%%%%%%%%%%%%%%%%%%%%%%%%%%%%%%%%%%%%%%%%%%%%%%%%%%%%%%%%%%%%%%%%%%%%%%%%

% properties of such compact remnants (e.g. mass and natal velocity) are still subject to an intense debate.
%Understanding stellar winds is tremendously important for the study of compact objects, because mass loss determines the pre-SN mass of a star (both its total mass and its core mass), which in turn affects the outcome of a SN explosion \cite{fryer1999,fryer2001,mapelli2009,belczynski2010}. 

Other aspects of massive star evolution also affect the pre-SN mass of a star. For example, surface magnetic fields appear to strongly quench stellar winds by magnetic confinement \cite{georgy2017,keszthelyi2017,petit2017}. %Figure~\ref{fig:petit} shows the equivalency curve between the reduction of mass loss due to metallicity and to magnetic confinement, as derived by \cite{petit2017}. 
In particular, \cite{petit2017} show that a non-magnetic star model with metallicity $\sim{}0.1$ Z$_\odot$ and a magnetic star model with solar metallicity and Alfv\'en radius $R_A\sim{}4\,{}R_\odot$ undergo approximately the same mass loss according to this model. This effect cannot be neglected because surface magnetic fields are detected in $\sim{}10$ per cent of the hot stars \cite{wade2016}, but is currently not included in models of compact-object formation.

%%%%%%%%%%%%%%%%%%%%%%%%%%%%%FIGURE %%%%%%%%%%%%%%%%%%%%%%%%%%%%%%%%%%%%%%%%
%\begin{figure}
%\center{
%\includegraphics[width=12cm]{petit2017_fig2}     % includes figure foo.eps
%}
%\caption{\label{fig:petit}Equivalency curve between the reduction of mass-loss due to metallicity (in units of $Z_\odot = 0.019$) and the reduction of mass-loss due to magnetic wind confinement (expressed as the extent of the Alfv\'en radius). The curve is coloured according to the mass-loss scaling. The metallicities of the Large Magellanic Cloud and the Small Magellanic Cloud, as well as the $\sim{}1/10$ Z$_\odot$ are indicated by horizontal red lines. The Alv\'en radius of a few known magnetic O-type stars are indicated with vertical blue lines. Figure~2 of \cite{petit2017}.}
%\end{figure}
%%%%%%%%%%%%%%%%%%%%%%%%%%%%%%%%%%%%%%%%%%%%%%%%%%%%%%%%%%%%%%%%%%%%%%%%%%%%

Finally,  rotation affects the evolution of a massive star in several ways (e.g. \cite{maeder2009,chieffi2013,limongi2017,limongi2018}). Describing the effects of rotation in detail is beyond the aims of this review. For this, we refer to the recent review of Marco Limongi \cite{limongi2017}.   As a general rule of thumb, rotation increases the stellar luminosity. This implies that mass loss is generally enhanced if rotation is accounted for. On the other hand, rotation also induces chemical mixing, which leads to the formation of larger Helium and Carbon-Oxygen cores. While enhanced mass loss implies smaller pre-SN masses, the formation of bigger cores has also strong implication for the final fate of a massive star, as we discuss in the following section.

\subsection{Supernovae (SNe)}
The mechanisms triggering iron core-collapse SNe are still highly uncertain. The basic framework and open issues are the following. As the mass of the central degenerate core reaches the Chandrasekhar mass \cite{chandrasekhar1931}, the degeneracy pressure of relativistic electrons becomes insufficient to support it against collapse. Moreover, electrons are increasingly removed, because protons capture them producing neutrons and neutrinos. This takes the core into a new state, where matter is essentially composed of neutrons, which support the core against collapse by their degeneracy pressure. To reach this new equilibrium, the core collapses from a radius of few thousand km down to a radius of few ten km in less than a second.  The gravitational energy gained from the collapse  is $W\sim{}5\times{}10^{53}\,{} {\rm erg} \,{}(m_{\rm PNS}/1.4\,{}M_\odot)^2\,{}(10\,{}{\rm km}/R_{\rm PNS})$, where $m_{\rm PNS}$ and $R_{\rm PNS}$ are the mass and radius of the proto-neutron star (PNS). 

The main problem is to explain how this gravitational energy can be (at least partially) transferred to the stellar envelope triggering the SN explosion \cite{colgatewhite1966,bethewilson1985}. 
Several mechanisms have been proposed, including rotationally-driven SNe and/or magnetically-driven SNe (see e.g. \cite{janka2012,fryer2014,foglizzo2015} and references therein). The most commonly investigated mechanism is the convective SN engine (see e.g. \cite{fryer2012}). According to this model, the collapsing core drives a bounce shock. For the SN explosion to occur, this shock must reverse the supersonic infall of matter from the outer layers of the star. Most of the energy in the shock consists in a flux of neutrinos. As soon as neutrinos are free to leak out (because the shock has become diffuse enough), their energy is lost and the shock stalls. The SN occurs only if the shock is revived by some mechanism. In the convective SN scenario, the region between the PNS surface and the shock stalling radius can become convectively unstable (e.g. because of a Rayleigh–Taylor instability). Such convective instability can  convert  the  energy  leaking out of the PNS in the form of neutrinos to kinetic energy pushing the convective region outward. If the convective region overcomes the ram pressure of the infalling material, the shock is revived and an explosion is launched. If not, the SN fails. %bounces, driving a shock. For the SN explosion to occur, this shock must reverse the supersonic infall of matter from the outer layers of the star. Most of the energy in the shock consists in a flux of neutrinos. As soon as neutrinos are free to leak out (because the shock has become diffuse enough), their energy is lost and the shock stalls. The SN occurs only if the shock is revived by some mechanism. If the region between the PNS surface and the shock stalling radius becomes convectively unstable (e.g. because of an entropy gradient), neutrinos can be injected more efficiently and deposit more energy into the shock, possibly reviving it. In other words, if the convective region overcomes the ram pressure of the infalling material, an explosion is launched. If not, the SN fails.

While this is the general idea of the convective engine, fully self-consistent simulations of core collapse with a state-of-the-art treatment of neutrino transport do not lead to explosions in spherical symmetry except for the lighter SN progenitors ($\lesssim{}10$ M$_\odot$, \cite{foglizzo2015,ertl2016}). Simulations which do not require the assumption of spherical symmetry (i.e. run at least in 2D) appear to produce successful explosions from first principles for a larger range of progenitor masses (see e.g. \cite{mullerjanka2012a,mullerjanka2012b}). However, 2D and 3D simulations  are still computationally challenging and cannot be used to make a study of the mass distribution of compact remnants.

Thus, in order to study compact-object masses, SN explosions are artificially induced by injecting in the pre-SN model some amount of kinetic energy (kinetic bomb) or thermal energy (thermal bomb) at an arbitrary mass location. The evolution of the shock is then followed by means of 1D hydrodynamical simulations with some relatively simplified treatment for neutrinos. This allows to simulate hundreds of stellar models.

Following this approach, O'Connor \& Ott (2011, \cite{oconnor2011}) propose a criterion to decide whether a SN is successful or not, based on the compactness parameter:
\begin{equation}\label{eq:compac}
\xi{}_m=\frac{m/{\rm M}_\odot}{R(m)/1000\,{}{\rm km}},
\end{equation}              
where $R(m)$ is the radius which encloses a given mass $m$. Usually, the compactness is defined for $m=2.5$ M$_\odot$ ($\xi_{2.5}$). \cite{oconnor2011} measure the compactness at core bounce\footnote{\cite{ugliano2012} show that $\xi_{2.5}$ is not significantly different at core bounce or at the onset of collapse.} in their simulations and find that the larger $\xi_{2.5}$ is, the shorter the time to form a BH (as shown in their Figure 6). This means that stars with a larger value of $\xi_{2.5}$ are more likely to collapse to a BH without SN explosion.

Rotating stellar models have lower values of $\xi_{2.5}$ (because of the centrifugal force) but produce lower neutrino luminosity and thus are more likely to form BHs without a SN explosion than non-rotating models \cite{oconnor2011}.

The work by Ugliano et al. (2012, \cite{ugliano2012}) and Horiuchi et al. (2014, \cite{horiuchi2014}) indicate that the best threshold between exploding and non-exploding models is $\xi_{2.5}\sim{}0.2$.

Finally \cite{ertl2016} stress that a single criterion (e.g. the compactness) cannot capture the complex physics of core-collapse SN explosions. They introduce a two-parameter criterion based on 
\begin{equation}
M_4=\frac{m(s=4)}{{\rm M}_\odot}\quad{}{\rm and}\quad{}\mu_4=\left[\frac{dm/{\rm M}_\odot}{dR/1000\,{}{\rm km}}\right]_{s=4},
\end{equation}
where $M_4$ is the mass (at the onset of collapse) where the dimensionless entropy per baryon is $s=4$, and $\mu_4$ is the spatial derivative at the location of $M_4$.

This choice is motivated by the fact that, in their 1D simulations, the explosion sets shortly after $M_4$ has fallen through the shock and well before the shell enclosing $M_4+0.3$ M$_\odot$ has collapsed. They show that exploding models can be distinguished from non-exploding models in the $\mu_4$ versus $M_4\,{}\mu_4$ plane (see their Figure 6) by a linear fit
\begin{equation}
y(x)=k_1\,{}x+k_2
\end{equation}
where $y(x)=\mu_4$, $x=M_4\,{}\mu_4$, and $k_1$ and $k_2$ are numerical coefficients which depend on the model (see Table 2 of \cite{ertl2016}). The reason of this behaviour is that $\mu{}_4$ scales with the rate of mass infall from the outer layers (thus the larger $\mu_4$ is, the lower the chance of the SN to occur), while $M_4\,{}\mu_4$ scales with the neutrino luminosity (thus the larger $M_4\,{}\mu_4$ is, the higher the chance of a SN explosion). Finally, \cite{ertl2016} stress that fallback is quite inefficient ($<0.05$ M$_\odot$) when the SN occur.

The models proposed by O'Connor \& Ott (2011, \cite{oconnor2011}) and Ertl et al. (2016, \cite{ertl2016}, see also \cite{sukhbold2014,sukhbold2016}) are sometimes referred to as the ``islands of explodability'' scenario, because they predict a non-monotonic behaviour of SN explosions with the stellar mass. This means, for example, that while a star with a mass $m=25$ M$_\odot$ and a star with a mass $m=29$ M$_\odot$ might end their life with a powerful SN explosion, another star with intermediate mass between these two (e.g. with a mass $m=27$ M$_\odot$) is expected to directly collapse to a BH without SN explosion. Thus, these models predict the existence of ``islands of explodability'', i.e. ranges of mass where a star is expected to explode, surrounded by mass intervals in which the star will end its life with a direct collapse.

The models discussed so far depend on quantities ($\xi_{2.5}$, $M_4$, $\mu_4$) which can be evaluated no earlier than the onset of core collapse. Thus, stellar evolution models are required which integrate a massive star till the iron core has formed. This is prohibitive for most stellar evolution models (with few remarkable exceptions, e.g. FRANEC \cite{chieffi2013} and MESA \cite{paxton2015}).

Fryer et al. (2012, \cite{fryer2012}) propose a simplified approach (see also \cite{fryer1999,fryer2001,fryer2006}). They suggest that the mass of the compact remnant depends mostly on two quantities: the Carbon-Oxygen core mass $m_{\rm CO}$ and the total final mass of the star $m_{\rm fin}$. In particular, $m_{\rm CO}$ determines whether the star will undergo a core-collapse SN or will collapse to a BH directly (namely, stars with $m_{\rm CO}>7.6$ M$_\odot$ collapse to a BH directly), whereas $m_{\rm fin}$ determines the amount of fallback on the proto NS. In this formalism, the only free parameter is the time to launch the shock. The explosion energy is significantly reduced if the shock is launched $\gg{}250$ ms after the onset of the collapse ({\it delayed} SN explosion) with respect to an explosion launched in the first $\sim{}250$ ms ({\it rapid} SN explosion, \cite{fryer2012}).

While this approach is quite simplified with respect to other prescriptions, \cite{limongi2017} and \cite{limongi2018} show that there is a strong correlation between the final Carbon-Oxygen mass and the compactness parameter $\xi{}_{2.5}$ at the onset of collapse, regardless of the rotation velocity of the progenitor star (see figure 21 of \cite{limongi2017}). Thus, we can conclude that the simplified models by \cite{fryer2012} can effectively describe the overall trend of a collapsing star, although they do not take into account several details of the stellar structure at the onset of collapse.

%Based on analytic calculations 
% (to obtain the pressure of the convective region and the ram pressure of the infalling material, see equations 1 and 2 of \cite{fryer2006}), \cite{fryer2012} estimate the maximum energy stored in the convective region.
%{\bf MM RIGUARDARE FRYER 1999, FRYER \& KALOGERA 2001}

%, they find a link between the mass of the Carbon-Oxygen core and the final fate of a SN explosion. 

% parlare di o'connor \& ott, ugliano,horiuchi, pejcha thompson?, ertl, clausen?

%parlare di fryer + limongi CO correlation

%fare una breve escursione nelle pair instability
Finally, it is important to recall pair-instability and pulsational pair-instability SNe \cite{fowler1964,barkat1967,rakavyshaviv1967,woosley2017}. If the Helium core of a star grows above $\sim{}30$ M$_\odot$ and the core temperature is  $\gtrsim{}7\times{}10^8$ K, the process of electron-positron pair production  becomes effective. It removes photon pressure from the core producing a sudden collapse before the iron core is formed.  For $m_{\rm He}>135$ M$_\odot$, the collapse cannot be reversed and the star collapses directly in to a BH \cite{woosley2017}. If $135\gtrsim{}m_{\rm He}\gtrsim{}64$ M$_\odot$, the collapse triggers an explosive burning of heavier elements, which has disruptive effects. This leads to a complete disruption of the star, leaving no remnant (the so-called pair-instability SN, \cite{hegerwoosley2002}).

For $64\gtrsim{}m_{\rm He}\gtrsim{}32$ M$_\odot$, pair production induces a series of pulsations of the core (pulsational pair instability SNe), which trigger an enhanced mass loss \cite{woosley2017}. At the end of this instability phase a remnant with non-zero mass is produced, significantly lighter than in case of a direct collapse. 

\subsection{The mass of compact remnants}\label{sec:remnants}
The previous sections suggest that our knowledge of the compact remnant mass is hampered by severe uncertainties, connected with both stellar winds and core-collapse SNe. Thus, models of the mass spectrum of compact remnants must be taken with a grain of salt. However, few robust features can be drawn. 
%%%%%%%%%%%%%%%%%%%%%%%%%%%%%FIGURE %%%%%%%%%%%%%%%%%%%%%%%%%%%%%%%%%%%%%%%%
\begin{figure}
\center{
\includegraphics[width=14cm]{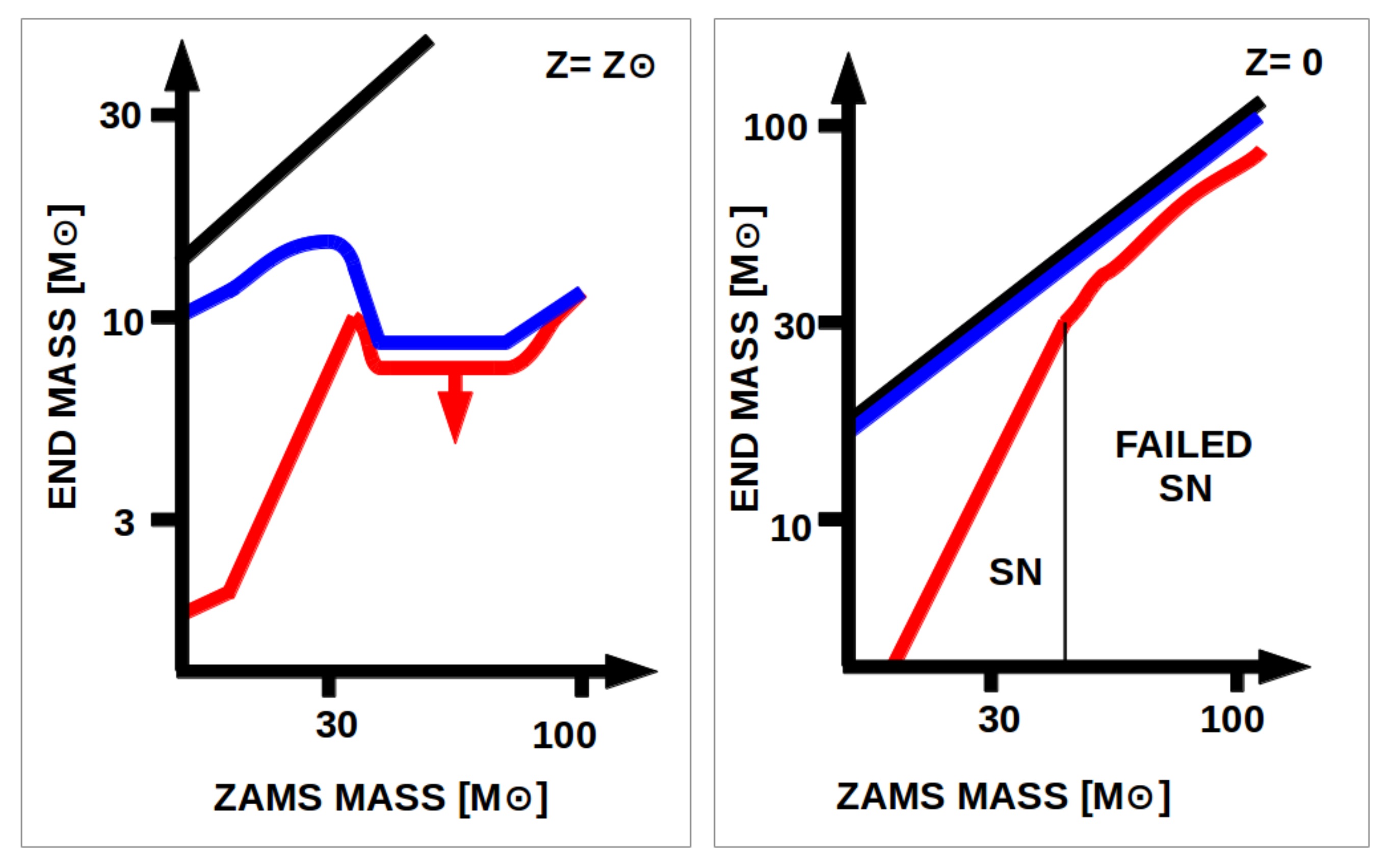}     % includes figure foo.eps
}
\caption{\label{fig:fromheger}Final mass of a star ($m_{\rm fin}$, blue lines) and mass of the compact remnant ($m_{rem}$, red lines) as a function of the ZAMS mass of the star. The thick black line marks the region where $m_{\rm fin}=m_{\rm ZAMS}$. Left-hand panel: solar metallicity star. Right-hand panel: metal-free star.  The red arrow on the left-hand panel is an upper limit for the remnant mass. Vertical thin black line in the right-hand panel: approximate separation between successful and failed SNe at $Z=0$. This cartoon was inspired by Figures 2 and 3 of Heger et al. (2003 \cite{heger2003}).}
\end{figure}
%%%%%%%%%%%%%%%%%%%%%%%%%%%%%%%%%%%%%%%%%%%%%%%%%%%%%%%%%%%%%%%%%%%%%%%%%%%%

Figure~\ref{fig:fromheger} is a simplified version of Figures 2 and 3 of Heger et al. (2003 \cite{heger2003}). The final mass of a star and the mass of the compact remnant are shown as a function of the ZAMS mass. The left and the right-hand panels show the case of a solar metallicity star and of a metal-free star, respectively. In the case of the solar metallicity star, the final mass of the star is much lower than the initial one, because stellar winds are extremely efficient. The mass of the compact remnant is also much lower than the final mass of the star because a core-collapse SN always takes place. 

In contrast, a metal-free star (i.e. a population~III star) loses a negligible fraction of its mass by stellar winds (the blue and the black line in Figure~\ref{fig:fromheger} are superimposed). As for the mass of the compact remnant, Figure~\ref{fig:fromheger} shows that there are two regimes: below a given threshold ($\approx{}30-40$ M$_\odot$) the SN explosion succeeds even at zero metallicity and the mass of the compact remnant is relatively small. Above this threshold, the mass of the star (in terms of both core mass and envelope mass) is sufficiently large that the SN fails. Most of the final stellar mass collapses to a BH, whose mass is significantly larger than in the case of a SN explosion.

What happens at intermediate metallicity between solar and zero, i.e. in the vast majority of the Universe we know? Predicting what happens to a metal-free star is relatively simple, because its evolution does not depend on the interplay between metals and stellar winds. The fate of a solar metallicity star is more problematic, because we must account for line-driven stellar winds, but most data we have about massive star winds are for nearly solar metallicity stars, which makes models easier to calibrate. Instead, modelling intermediate metallicities is significantly more complicated, because the details depend on the interplay between metals and stellar winds and only limited data are available for calibration (mostly data for the Large and Small Magellanic Clouds).

As a rule of thumb (see e.g. \cite{fryer2012,spera2015}), we can draw the following considerations. If the zero-age main sequence (ZAMS) mass of a star is large ($m_{\rm ZAMS}\gtrsim{}30$ M$_\odot$), then the amount of mass lost by stellar winds is the main effect which determines the mass of the compact remnant. At low metallicity ($\lesssim{}0.1$ Z$_\odot$) and for a low Eddington factor ($\Gamma_e<0.6$), mass loss by stellar winds is not particularly large. Thus, the final mass $m_{\rm fin}$ and the Carbon-Oxygen mass $m_{\rm CO}$ of the star may be sufficiently large to avoid a core-collapse SN explosion: the star may form a massive BH ($\gtrsim{}20$ M$_\odot$) by direct collapse, unless a pair-instability or a pulsational-pair instability SN occurs. At high metallicity ($\approx{}$Z$_\odot$) or large Eddington factor ($\Gamma_e>0.6$), mass loss by stellar winds is particularly efficient and may lead to a small $m_{\rm fin}$ and $m_{\rm CO}$: the star is expected to undergo a core-collapse SN and to leave a relatively small remnant. 

If the ZAMS mass of a star is relatively low ($7<m_{\rm ZAMS}<30$ M$_\odot$), then stellar winds are not important (with the exception of super asymptotic giant branch stars), regardless of the metallicity. In this case, the details of the SN explosion (e.g. energy of the explosion and amount of fallback) are crucial to determine the final mass of the remnant.
 
This general sketch may be affected by several factors, such as pair-instability SNe, pulsational pair-instability SNe (e.g. \cite{woosley2017}) and an {\it island scenario} for core-collapse SNe (e.g. \cite{ertl2016}).

\begin{figure}
\center{
\includegraphics[width=13cm]{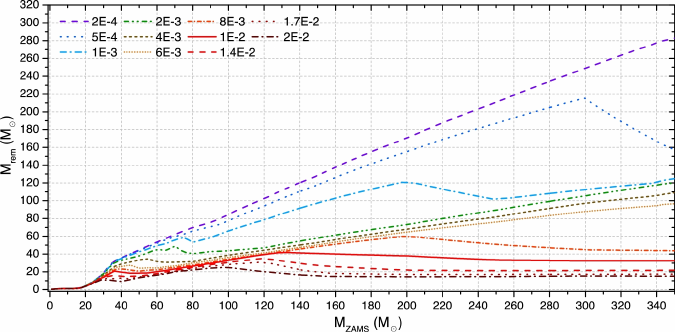}     % includes figure foo.eps
\includegraphics[width=13.2cm]{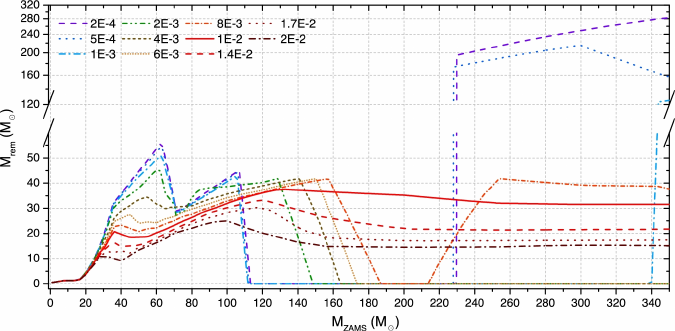}     % includes figure foo.eps
}
\caption{\label{fig:spera2017}Mass of the compact remnant ($m_{\rm rem}$) as a function of the ZAMS mass of the star ($m_{\rm ZAMS}$). Lower (upper) panel: pulsational pair-instability and pair-instability SNe are (are not) included. In both panels: dash-dotted brown line: $Z = 2.0\times{} 10^{-2}$; dotted dark orange line: $Z = 1.7\times{}10^{-2}$; dashed red line: $Z = 1.4\times{}10^{-2}$; solid red line: $Z = 1.0\times{} 10^{-2}$; short dash-dotted orange line: $Z = 8.0 \times{} 10^{-3}$; short dotted light orange line: $Z = 6.0 \times{} 10^{-3}$; short dashed green line: $Z = 4.0 \times{}10^{-3}$; dash-double dotted green line: $Z = 2.0 \times{} 10^{-3}$; dash-dotted light blue line: $Z = 1.0 \times{} 10^{-3}$; dotted blue line: $Z = 5.0\times{} 10^{-4}$; dashed violet line: $Z = 2.0\times{} 10^{-4}$. A delayed core-collapse SN mechanism has been assumed, following the prescriptions of \cite{fryer2012}. This Figure was adapted from Figures~1 and 2 of Spera \&{} Mapelli (2017, \cite{spera2017}).}
\end{figure}

The effect of pair-instability and pulsational pair-instability SNe is clearly shown in Figure~\ref{fig:spera2017}. The top panel was obtained accounting only for stellar evolution and core-collapse SNe. In contrast, the bottom panel also includes pair-instability and pulsational pair-instability SNe. This figure shows that the mass of the compact remnant strongly depends on the metallicity of the progenitor star if $m_{\rm ZAMS}\gtrsim{}30$ M$_\odot$. In most cases, the lower the metallicity of the progenitor is, the larger the maximum mass of the compact remnant \cite{heger2003,mapelli2009,belczynski2010,mapelli2010,mapelli2013,spera2015,spera2017}. However, for metal-poor stars  ($Z<10^{-3}$) with ZAMS mass $230>m_{\rm ZAMS}>110$ M$_\odot$ pair instability SNe lead to the complete disruption of the star and no remnant is left. Only very massive ($m_{\rm ZAMS}>230$ M$_\odot$) metal-poor  ($Z<10^{-3}$) stars can collapse to a BH directly, producing intermediate-mass BHs (i.e, BHs with mass $\gtrsim{}100$ M$_\odot$).

If $Z<10^{-3}$ and  $110>m_{\rm ZAMS}\gtrsim{}60$ M$_\odot$, the star enters the pulsational pair-instability SN regime: mass loss is enhanced and the final BH mass is smaller ($m_{\rm BH}\sim{}30-55$ M$_\odot$, bottom panel of Fig.~\ref{fig:spera2017}) than we would have expected from direct collapse ($m_{\rm BH}\sim{}50-100$ M$_\odot$, top panel of Fig.~\ref{fig:spera2017}). Thus, accounting for both pair instability and pulsational pair-instability SNe leads to a {\emph{BH mass gap}}\footnote{The existence of a BH mass gap between $\sim{}50$ and $\sim{}100$ M$_\odot$ is currently consistent with GW detections (see e.g. \cite{fishbach2017,talbot2018,wysocki2018}).} between $m_{\rm BH}\sim{}60$ M$_\odot$ and $m_{\rm BH}\sim{}120$ M$_\odot$.

Finally, the mass spectrum of relatively low-mass stars ($8<m_{\rm ZAMS}<30$ M$_\odot$) is not significantly affected by metallicity. The assumed core-collapse SN model is the most important factor in this mass range \cite{fryer2012}.

%putting everything together: spera and mapelli...
%%%%%%%%%%%%%%%%%%%%%%%%%%%%%FIGURE %%%%%%%%%%%%%%%%%%%%%%%%%%%%%%%%%%%%%%%%
%%%%%%%%%%%%%%%%%%%%%%%%%%%%%%%%%%%%%%%%%%%%%%%%%%%%%%%%%%%%%%%%%%%%%%%%%%%%

\subsection{Natal kicks}
Compact objects are expected to receive a natal kick from the parent SN explosion, because of asymmetries in the neutrino flux and/or in the ejecta (see \cite{janka2012} for a review). 
%The natal kick should arise from asymmetries during the SN explosion (e.g. asymmetries in neutrino emission and/or in the ejection of the outer layers of the parent star).
The natal kick has a crucial effect on the evolution of a BHB, because it can either unbind the binary or change its orbital properties. For example, a SN kick can increase the orbital eccentricity or misalign the spins of the two members of the binary.

Unfortunately, it is extremely difficult to quantify natal kicks from state-of-the-art SN simulations and measurements of natal kicks are scanty, especially for BHs.

As to NSs, indirect observational estimates of SN kicks give contrasting results. Hobbs et al. (2005, \cite{hobbs2005}) found that a single Maxwellian with root mean square $\sigma{}_{\rm CCSN}=265$ km s$^{-1}$ can match the proper motions of 233 single pulsars in the Milky Way. Other works suggest a bimodal velocity distribution, with a first peak at low velocities (e.g. $\sim{}0$ km s$^{-1}$ according to \cite{fryer1998} or $\sim{}90$ km s$^{-1}$ according to \cite{arzoumanian2002}) and a second peak at high velocities ($>600$ km s$^{-1}$ according to \cite{fryer1998} or $\sim{}500$ km s$^{-1}$ for \cite{arzoumanian2002}). Similarly, the recent work of Verbunt et al. (2017, \cite{verbunt2017}) indicates that a double Maxwellian distribution provides a significantly better fit to the observed velocity distribution than a single Maxwellian. Finally, the analysis of Beniamini \&{} Piran (2016, \cite{beniamini2016}) shows that low kick velocities ($\lesssim{}30$ km s$^{-1}$) are required to match the majority of Galactic DNSs, especially those with low eccentricity.

A possible interpretation of these observational results is that natal kicks depend on the SN mechanism (e.g. electron-capture versus core-collapse SN, e.g. \cite{giacobbo2018c}) or on the binarity of the NS progenitor. For example, if the NS progenitor evolves in a close binary system (i.e. in a binary system where the two stars have exchanged mass with each other, see Section~\ref{sec:masstransfer}) it might undergo an ultra-stripped SN (see \cite{tauris2017} and references therein for more details). A star can undergo an ultra-stripped SN explosion only if it was heavily stripped by mass transfer to a companion \cite{tauris2013,tauris2015}.  
The natal kick of an ultra-stripped SN should be low \cite{tauris2015}, because of the small mass of the ejecta ($\lesssim{}0.1$ M$_\odot$). Low kicks ($\lesssim{}50$ km s$^{-1}$) for ultra-stripped core-collapse SNe are also confirmed by recent hydrodynamical simulations \cite{suwa2015,janka2017}.

%According to this interpretation, NSs born in close binaries might have substantially smaller kicks than NSs born from single stars or from stars in detached binaries.

As to BHs, the only indirect measurements of natal kicks arise from spatial distributions, proper motions and orbital properties of BHs in X-ray binaries (e.g. \cite{mirabel2017}). Evidence for a relatively small natal kick has been found for both GRO~J1655--40 \cite{willems2005} and Cygnus X-1 \cite{wong2012}, whereas H~1705--250 \cite{repetto2012,repetto2017} and XTE~J$1118+480$ \cite{mirabel2001,fragos2009} require high kicks  ($>100$ km s$^{-1}$). By analysing the position of BHs in X-ray binaries with respect to the Galactic plane, Repetto et al. (2012, \cite{repetto2012}) suggest that BH natal kicks should be as high as NS kicks. Recently, Repetto et al. (2017, \cite{repetto2017}) perform a similar analysis but accounting also for binary evolution, and find that at least two BHs in X-ray binaries (H~1705--250 and XTE~J$1118+480$) require high kicks.

Most models of BHB evolution assume that natal kicks of BHs are drawn from the same distribution as NS kicks, but reduced by some factor. For example, linear momentum conservation suggests that
\begin{equation}
  v_{\rm BH}=\frac{m_{\rm NS}}{m_{\rm BH}}\,{}v_{\rm NS},
\end{equation}
where $v_{\rm BH}$ is the natal kick of a BH with mass $m_{\rm BH}$ and $v_{\rm NS}$ is the natal kick of a NS with mass $m_{\rm NS}$.

Alternatively, the natal kick can be reduced by the amount of fallback, under the reasonable assumption that fallback quenches the initial asymmetries. Following \cite{fryer2012},
\begin{equation}
  v_{\rm BH}=(1-f_{\rm fb})\,{}v_{\rm NS},
\end{equation}
where $f_{\rm fb}$ quantifies the fallback ($f_{\rm fb}=0$ for no fallback and $f_{\rm fb}=1$ for direct collapse).

Finally, most studies assume that  BHs born from direct collapse receive no kick \cite{fryer2012}. In summary, natal kicks are one of the most debated issues about compact objects. Their actual amount has dramatic implications on the merger rate and on the properties (spin and mass distribution) of merging compact objects.

\section{Binaries of stellar black holes}
Naively, one could think that if two massive stars are members of a binary system, they will eventually become a double BH binary and the mass of each BH will be the same as if its progenitor star was a single star. This is true only if the binary system is sufficiently wide (detached binary) for its entire evolution. If the binary is close enough, it will evolve through several processes which might significantly change its final fate. %Here, we will  mention some of the most important ones.

The so-called binary population-synthesis codes have been used to investigate the effect of binary evolution processes on the formation of BHBs in isolated binaries (e.g. \cite{portegieszwart1996,hurley2002,podsiadlowski2003,belczynski2008,mapelli2013,mennekens2014,eldridge2016,mapellietal2017,stevenson2017,giacobbo2018,barrett2018,giacobbo2018b,giacobbo2018c,kruckow2018,eldridge2018,spera2018}). These are Monte-Carlo based codes which combine a description of stellar evolution with prescriptions for supernova explosions and with a formalism for binary evolution processes.

In the following, we mention some of the most important binary-evolution processes and we briefly discuss their treatment in the most used population-synthesis codes.

\subsection{Mass transfer}\label{sec:masstransfer}
If two stars exchange matter to each other, it means they undergo a mass transfer episode. This might be driven either by stellar winds or by an episode of Roche-lobe filling.

When a massive star loses mass by stellar winds, its companion might be able to capture some of this mass. This will depend on the amount of mass which is lost and on the relative velocity of the wind with respect to the companion star. Based on the Bondi \& Hoyle (1944, \cite{bondi1944}) formalism, Hurley et al. (2002, \cite{hurley2002}) describe the mean mass accretion rate by stellar winds as
\begin{equation}
\dot{m}_2=\frac{1}{\sqrt{1-e^2}}\,{}\left(\frac{G\,{}m_2}{v_{\rm w}^2}\right)^2\,{}\frac{\alpha_{\rm w}}{2\,{}a^2}\,{}\frac{1}{[1+(v_{\rm orb}/v_{\rm w})^2]^{3/2}}\,{}\dot{m}_{\rm 1},
\end{equation}
where $e$ is the binary eccentricity, $G$ is the gravitational constant, $m_2$ is the mass of the accreting star, $v_{\rm w}$ is the velocity of the wind, $\alpha_{\rm w}\sim{}3/2$ is an efficiency constant, $a$ is the semi-major axis of the binary, $v_{\rm orb}=\sqrt{G\,{}(m_1+m_2)/a}$ is the orbital velocity of the binary ($m_1$ being the mass of the donor), and $\dot{m}_1$ is the mass loss rate by the donor (here we assume $\dot{m}_1\ge{}0$). Since $\dot{m}_1$ is usually quite low ($<10^{-3}$ M$_\odot$ yr$^{-1}$) and $v_{\rm w}$ is usually quite high  ($>1000$ km s$^{-1}$ for a line-driven wind) with respect to the orbital velocity, this kind of mass transfer is usually rather inefficient.

Mass transfer by Roche lobe overflow is usually more efficient. The Roche lobe of a star in a binary system is the maximum equipotential surface around the star within which matter is bound to the star. While the exact shape of the Roche lobe should be calculated numerically, a widely used approximate formula \cite{eggleton1983} is
\begin{equation}\label{eq:rlobe}
r_{\rm L,1}=a\,{}\frac{0.49\,{}q^{2/3}}{0.6\,{}q^{2/3}+\ln{\left(1+q^{1/3}\right)}},
\end{equation}
where $a$ is the semi-major axis of the binary and $q=m_1/m_2$ ($m_1$ and $m_2$ are the masses of the two stars in the binary). This formula describes the Roche lobe of star with mass $m_1$, while the corresponding Roche lobe of star with mass $m_2$ ($r_{\rm L,2}$) is obtained by swapping the indexes.

The Roche lobes of the two stars in a binary are thus connected by the L1 Lagrangian point. Since the Roche lobes are equipotential surfaces, matter orbiting at or beyond the Roche lobe can flow freely from one star to the other. We say that a star overfills (underfills) its Roche lobe when its radius is larger (smaller) than the Roche lobe. If a star overfills its Roche lobe, a part of its mass flows toward the companion star which can accrete (a part of) it. The former and the latter are thus called donor and accretor star, respectively. 

Mass transfer obviously changes the mass of the two stars in a binary, and thus the final mass of the compact remnants of such stars, but also the orbital properties of the binary. If mass transfer is non conservative (which is the most realistic case in both mass transfer by stellar winds and Roche lobe overflow), it leads to an angular momentum loss, which in turn affects the semi-major axis.

An important point about Roche lobe overflow is to estimate whether it is (un)stable and on which timescale. A commonly used approach consists in comparing the following quantities \cite{webbink1985,portegieszwart1996,tout1997,hurley2002}:
\begin{eqnarray}
\zeta_{\rm ad}=\left(\frac{d\ln{R_1}}{d\ln{m_1}}\right)_{\rm ad}\nonumber\\
\zeta_{\rm th}=\left(\frac{d\ln{R_1}}{d\ln{m_1}}\right)_{\rm th}\nonumber\\
\zeta_{\rm L}=\left(\frac{d\ln{r_{\rm L,1}}}{d\ln{m_1}}\right),
\end{eqnarray}
where $\zeta_{\rm ad}$ is the change of radius of the donor (induced by the mass loss) needed to adiabatically adjust the star to a new hydrostatic equilibrium, $\zeta_{\rm th}$ is the change of the radius of the donor (induced by the mass loss) needed to adjust the star to a new thermal equilibrium, and $\zeta_{\rm L}$ is the change of the Roche lobe of the donor (induced by the mass loss).

If $\zeta_{\rm L}>\zeta_{\rm ad}$, then the star expands faster than the Roche lobe (on conservative mass transfer) and mass transfer is dynamically unstable. If $\zeta_{\rm ad}>\zeta_{\rm L}>\zeta_{\rm th}$, then mass transfer becomes unstable over a Kelvin-Helmholtz timescale. Finally, if ${\rm min}(\zeta_{\rm ad},\zeta_{\rm th})>\zeta_{\rm L}$, mass transfer is stable until nuclear evolution causes a further expansion (or contraction) of the radius.

If mass transfer is dynamically unstable ($\zeta_{\rm L}>\zeta_{\rm ad}$) or both stars overfill their Roche lobe, then the binary is expected to merge -- if the donor lacks a steep density gradient between the core and the envelope --, or to enter common envelope (CE) -- if the donor has a clear distinction between core and envelope.

\subsection{Common envelope (CE)}
If two stars enter in CE, their envelope(s) stop co-rotating with their cores. The two stellar cores (or the compact remnant and the core of the star, if the binary is already single degenerate) are embedded in the same non-corotating envelope and start spiralling in as an effect of gas drag exerted by the envelope. Part of the orbital energy lost by the cores as an effect of this drag is likely converted into heating of the envelope, making it more loosely bound. If this process leads to the ejection of the envelope, then the binary survives, but the post-CE binary is composed of two naked stellar cores (or a compact remnant and a naked stellar core). Moreover, the orbital separation of the two cores (or the orbital separation of the compact remnant and the core) is considerably smaller than the initial orbital separation of the binary, as an effect of the spiral in\footnote{Note that a short-period (from few hours to few days) binary system composed of a naked Helium core and BH might be observed as an X-ray binary, typically a Wolf-Rayet X-ray binary. In the local Universe, we know a few ($\sim{}7$) Wolf-Rayet X-ray binaries, in which a compact object (BH or NS) accretes mass through the wind of the naked stellar companion (see e.g. \cite{esposito2015} for more details). These rare X-ray binaries are thought to be good progenitors of merging compact-object binaries.}. This circumstance is crucial for the fate of a BH binary. In fact, if the binary which survives a CE phase evolves into a double BH binary, this double BH binary will have a very short semi-major axis, much shorter than the sum of the maximum radii of the progenitor stars, and may be able to merge by GW emission within a Hubble time.

\begin{figure}
\center{
\includegraphics[width=14cm]{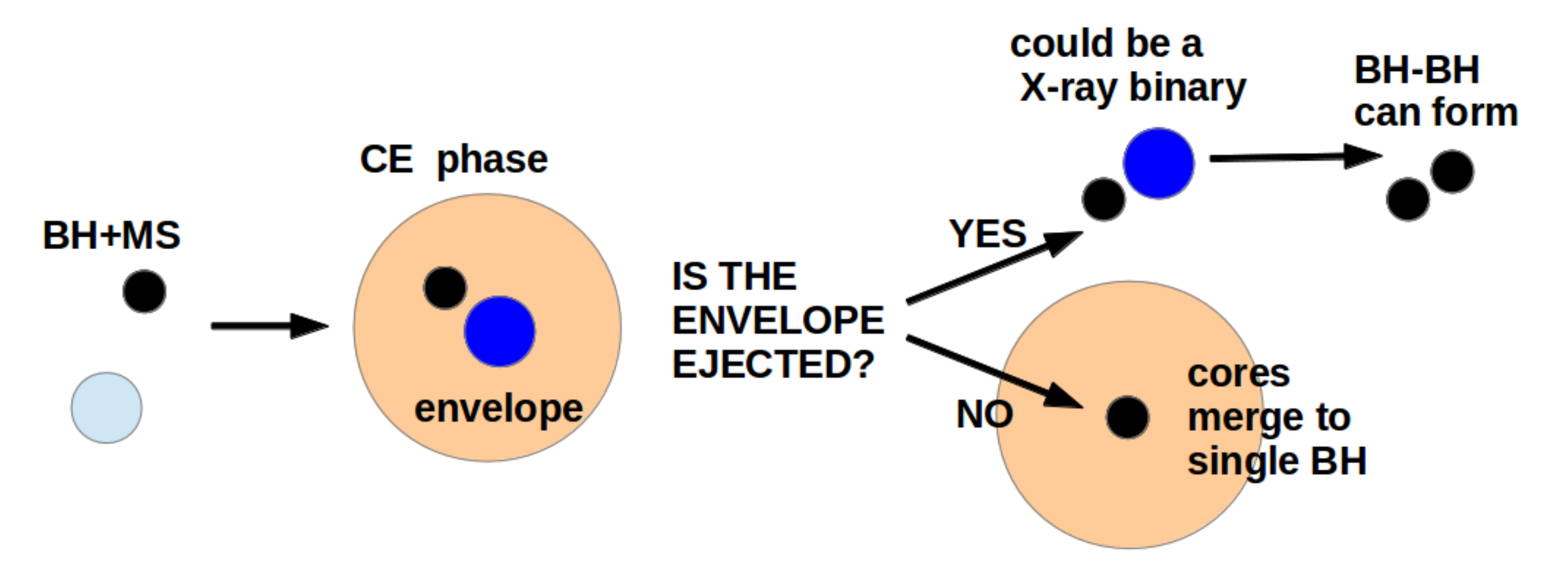}     % includes figure foo.eps
}
\caption{\label{fig:commonenv}Cartoon of the evolution of a BH binary which evolves through a CE phase. The companion of the BH is initially in the main sequence (MS). In the cartoon, the BH is indicated by a black circle, while the MS companion is indicated by the light blue circle. When the companion evolves off the MS, becoming a giant star, it overfills its Roche lobe. The BH and the giant star enter a CE (the CE is indicated in orange, while the core of the giant is represented by the dark blue circle). The core of the giant and the BH spiral in because of the gas drag exerted by the envelope. If the envelope is ejected, we are left with a new binary, composed of the BH and the naked Helium core of the giant. The new binary has a much smaller orbital separation than the initial binary. If the naked Helium core becomes a BH and its natal kick does not disrupt the binary, then a double BH binary is born, possibly with a small semi-major axis. In contrast, if the envelope is not ejected, the BH and the Helium core spiral in, till they merge together. A single BH is left, which will not be a source of GWs.}
\end{figure}

In contrast, if the envelope is not ejected, the two cores (or the compact remnant and the core) spiral in till they eventually merge. This premature merger of a binary during a CE phase prevents the binary from evolving into a double BH binary. The cartoon in Figure~\ref{fig:commonenv} summarizes these possible outcomes.
%%%%%%%%%%%%%%%%%%%%%%%%%%%%%FIGURE %%%%%%%%%%%%%%%%%%%%%%%%%%%%%%%%%%%%%%%%
%%%%%%%%%%%%%%%%%%%%%%%%%%%%%%%%%%%%%%%%%%%%%%%%%%%%%%%%%%%%%%%%%%%%%%%%%%%%

The $\alpha{}\lambda{}$ formalism \cite{webbink1984} is the most common formalism adopted to describe a common envelope. The basic idea of this formalism is that the energy needed to unbind the envelope comes uniquely from the loss of orbital energy of the two cores during the spiral in.

The fraction of the orbital energy of the two cores which goes into unbinding the envelope can be expressed as
\begin{equation}
\Delta{}E=\alpha{}\,{}(E_{\rm b,f}-E_{\rm b,i})=\alpha{}\,{}\frac{G\,{}m_{\rm c1}\,{}m_{\rm c2}}{2}\,{}\left(\frac{1}{a_{\rm f}}-\frac{1}{a_{\rm i}}\right),
\end{equation}
where $E_{\rm b,i}$ ($E_{\rm b,f}$) is the orbital binding energy of the two cores before (after) the CE phase, $a_{\rm i}$ ($a_{\rm f}$) is the semi-major axis before (after) the CE phase, $m_{\rm c1}$ and $m_{\rm c2}$ are the masses of the two cores, and $\alpha{}$ is a dimensionless parameter that measures which fraction of the removed orbital energy is transferred to the envelope. If the primary is already a compact object (as in Figure~\ref{fig:commonenv}), $m_{\rm c2}$ is the mass of the compact object.

The binding energy of the envelope is
\begin{equation}
E_{\rm env}=\frac{G}{\lambda}\,{}\left[\frac{m_{\rm env,1}\,{}m_{\rm 1}}{R_1}+\frac{m_{\rm env,2}\,{}m_{\rm 2}}{R_2}\right],
\end{equation}
where $m_1$ and $m_2$ are the masses of the primary and the secondary member of the binary, $m_{\rm env,1}$ and $m_{\rm env,2}$ are the masses of the envelope of the primary and the secondary member of the binary, $R_1$ and $R_2$ are the radii of the primary and the secondary member of the binary, and $\lambda{}$ is the parameter which measures the concentration of the envelope (the smaller $\lambda{}$ is, the more concentrated is the envelope).

By imposing $\Delta{}E=E_{\rm env}$ we can derive what is the value of the final semi-major axis $a_{\rm f}$ for which the envelope is ejected:
\begin{equation}\label{eq:CE}
\frac{1}{a_{\rm f}}=\frac{1}{\alpha{}\,{}\lambda{}}\,{}\frac{2}{m_{\rm c1}\,{}m_{\rm c2}}\,{}\left[\frac{m_{\rm env,1}\,{}m_{\rm 1}}{R_1}+\frac{m_{\rm env,2}\,{}m_{\rm 2}}{R_2}\right]+\frac{1}{a_{\rm i}}.
\end{equation}
If $a_{\rm f}$ is lower than the sum of the radii of the two cores (or than the sum of the Roche lobe radii of the cores), then the binary will merge during CE, otherwise the binary survives and equation~\ref{eq:CE} tells us the final orbital separation. This means that the larger (smaller) $\alpha{}\,{}\lambda{}$ is, the larger (smaller) the final orbital separation.

Actually, we have known for a long time (see \cite{ivanova2013} for a review) that this simple formalism is a poor description of the physics of CE, which is considerably more complicated. For example, there is a number of observed systems for which an $\alpha{}>1$ is required, which is obviously un-physical. Moreover, $\lambda{}$ cannot be the same for all stars. It is expected to vary wildly not only from star to star but also during different evolutionary stages of the same star. Several authors \cite{xu2010,loveridge2011} have estimated $E_{\rm env}$ directly from their stellar models, which removes the $\lambda{}$ parameter from equation~\ref{eq:CE} and significantly improves this formalism. However, even in this case, we cannot get rid of the $\alpha{}$ parameter. 

\begin{figure}
\center{
\includegraphics[width=12cm]{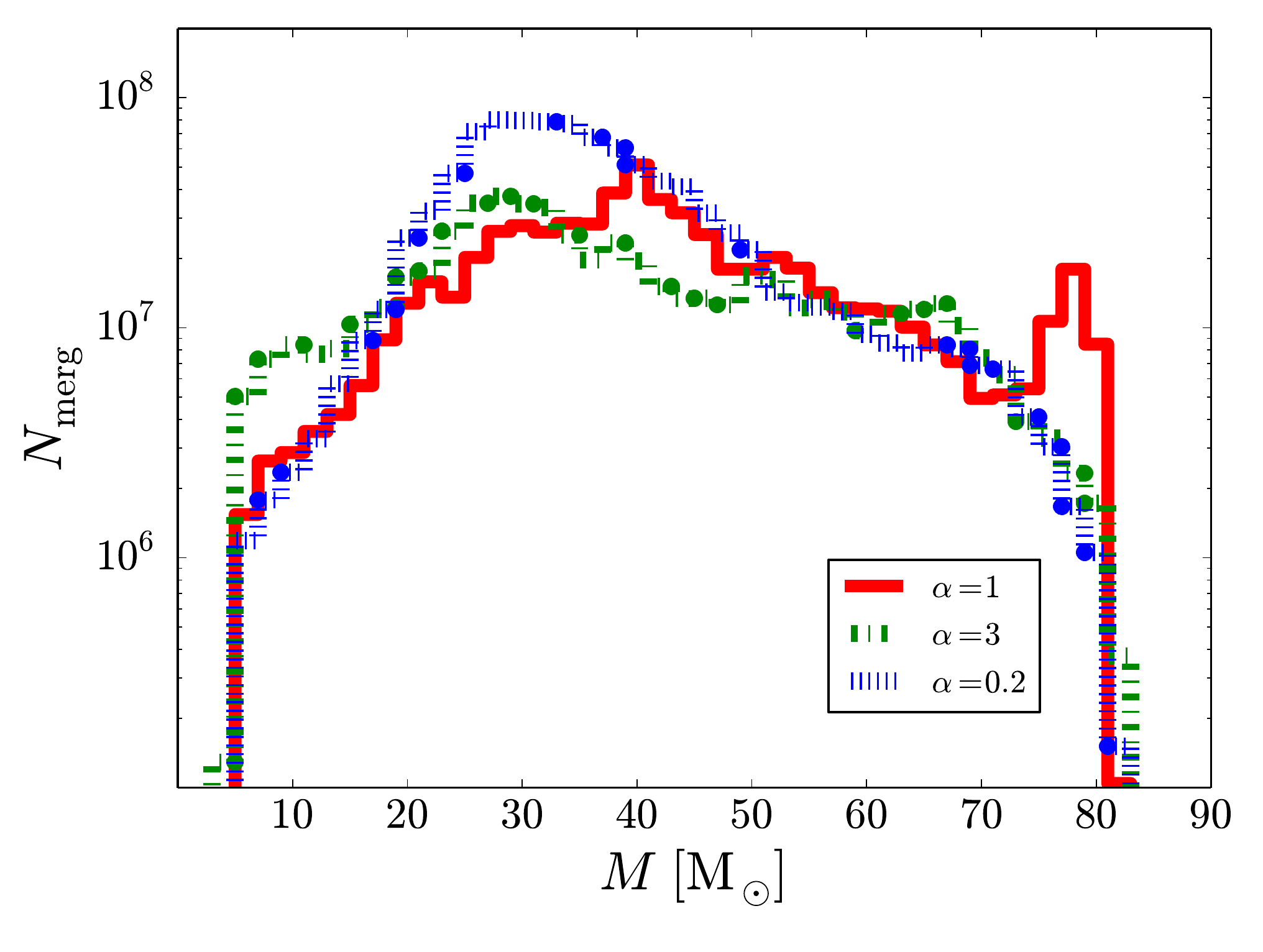}     % includes figure foo.eps
}
\caption{\label{fig:CEmobse}Distribution of total masses ($M=m_1+m_2$) of merging BH binaries in the LIGO-Virgo instrumental horizon, obtained with the {\sc mobse} code \cite{giacobbo2018,giacobbo2018b,giacobbo2018c}. Merging binaries come from progenitors with different metallicity and were sampled as described in \cite{mapellietal2017} (see also \cite{mapelli2018}). The only difference between the three histograms is the value of $\alpha{}$ in the CE formalism. Red solid line: $\alpha{}=1$; green dot-dashed line: $\alpha{}=3$; blue dotted line: $\alpha{}=0.2$.}
\end{figure}

Figure~\ref{fig:CEmobse} shows the distribution of total masses of merging BHs obtained with the same code ({\sc MOBSE}, \cite{giacobbo2018}) by changing solely the value of $\alpha{}$ (while $\lambda{}$ is calculated self-consistently by the code as described in \cite{claeys2014}). The difference between the three mass distributions is a clear example of how important is CE for the demography of BH binaries.
%%%%%%%%%%%%%%%%%%%%%%%%%%%%%FIGURE %%%%%%%%%%%%%%%%%%%%%%%%%%%%%%%%%%%%%%%%
%%%%%%%%%%%%%%%%%%%%%%%%%%%%%%%%%%%%%%%%%%%%%%%%%%%%%%%%%%%%%%%%%%%%%%%%%%%%

Thus, it would be extremely important to model the CE in detail, for example with numerical simulations. A lot of effort has been put on this in the last few years, but there are still many open questions (see the review by \cite{ivanova2013}). For example, we do not have self-consistent models of the onset of CE, when an unstable mass transfer prevents the envelope from co-rotating with the core. Usually, hydrodynamical simulations of CE start when the core of the companion is already at the surface of the envelope.

The only part of CE which has been successfully modelled by several authors (e.g. \cite{rickertaam2008,rickertaam2012,passy2012,ohlmann2016}) is the initial spiral in phase, when the two cores spiral in on a dynamical time scale ($\approx{}100$ days). 

However, at the end of this dynamical spiral in only a small fraction of the envelope ($\sim{}25$ per cent, \cite{ohlmann2016}) appears to be ejected in most simulations. When the two cores are sufficiently close that they are separated only by a small gas mass, the spiral in slows down and the system evolves on the Kelvin-Helmholtz timescale of the envelope ($\approx{}10^{3-5}$ years). Simulating the system for a Kelvin-Helmholtz timescale is prohibitive for current simulations. Thus, the description of CE remains the conundrum of massive binary evolution.

%connected to an equipotential surface around the companion star by 

%descrivere in dettaglio common envelope e roche lobe
%{\bf menzionare i tides che possono allineare gli spin e i SN kicks?}

\subsection{Alternative evolution to CE}
%{\bf parlare del chemical mixing scenario}
Massive fast rotating stars can have a chemically homogeneous evolution (CHE): they do not develop a chemical composition gradient because of the mixing induced by rotation. This is particularly true if the star is metal poor, because stellar winds are not efficient in removing angular momentum. If a binary is very close, the spins of its members are even increased during stellar life, because of tidal synchronisation. The radii of stars following CHE are usually much smaller than the radii of stars developing a chemical composition gradient \cite{demink2016,mandel2016}. This implies that even very close binaries (few tens of solar radii) can avoid CE.

Marchant et al. (2016, \cite{marchant2016}) simulate very close binaries whose components are fast rotating massive stars. A number of their simulated binaries evolve into contact binaries where both binary components fill and even overfill their Roche volumes. If metallicity is sufficiently low and rotation sufficiently fast, these binaries may evolve as ``over-contact'' binaries: the over-contact phase differs from a classical CE phase because co-rotation can, in principle, be maintained as long as material does not overflow the L2 point. This means that a spiral-in that is due to viscous drag can be avoided, resulting in a stable system evolving on a nuclear timescale.

Such over-contact binaries maintain relatively small stellar radii during their evolution (few ten solar radii) and may evolve into a double BH binary with a very short orbital period. This scenario predicts the formation of merging BHs with relatively large masses ($>20$ M$_\odot$), nearly equal mass ($q=1$), and with large aligned spins. The latter prediction is quite at odds with the effective spins of GW150914, GW170104, and GW170814 (see Table~\ref{tab:table1}).

\subsection{Summary of the isolated binary formation channel}
In this section, we have highlighted the most important aspects and the open issues of the ``isolated binary formation scenario'', i.e.  the model which predicts the formation of merging BHs through the evolution of isolated binaries. For isolated binaries we mean stellar binary systems which are not perturbed by other stars or compact objects.

%%%%%%%%%%%%%%%%%%%%%%%%%%%%%FIGURE %%%%%%%%%%%%%%%%%%%%%%%%%%%%%%%%%%%%%%%%
\begin{figure}
\center{
\includegraphics[width=12cm]{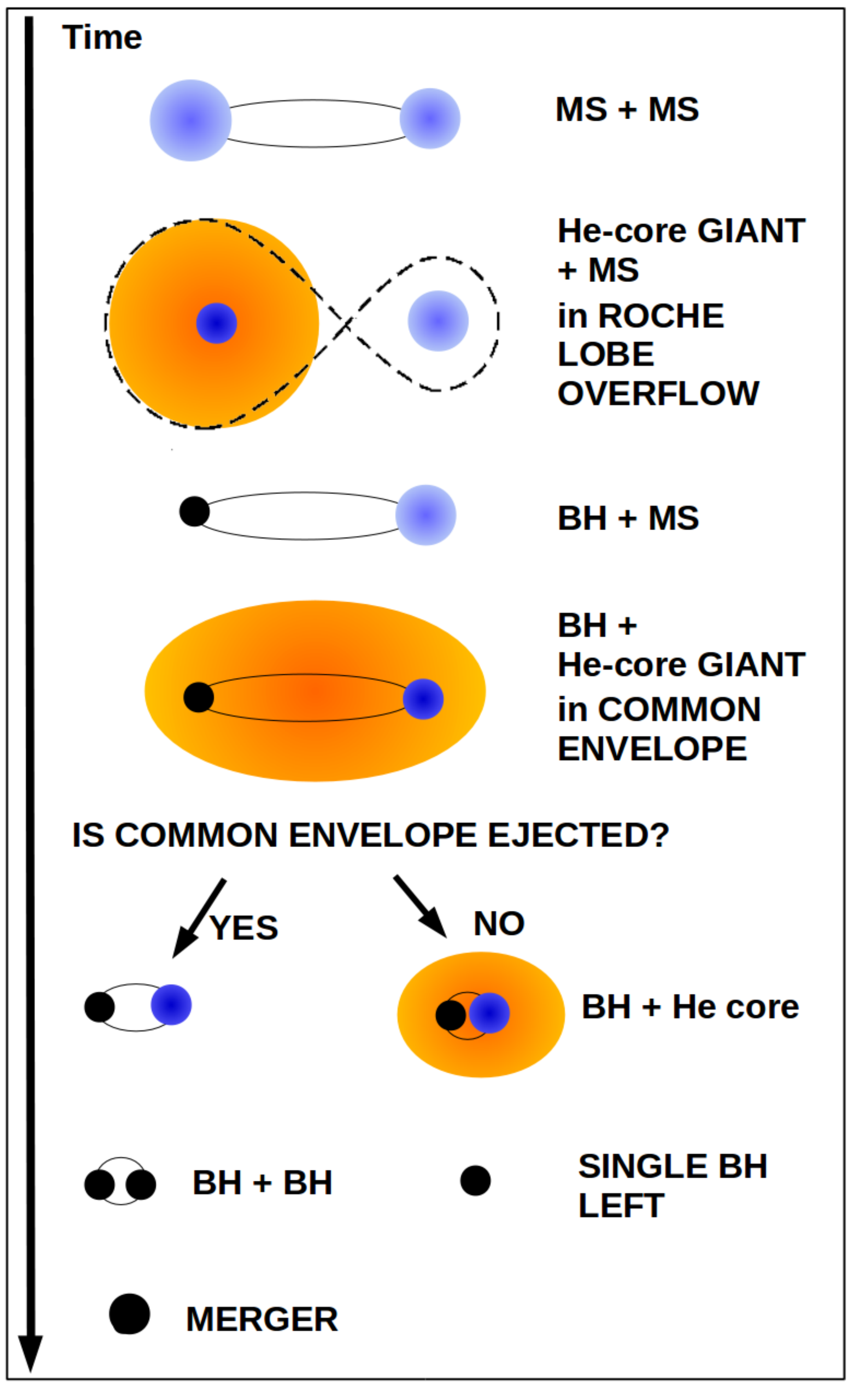}     % includes figure foo.eps
}
\caption{\label{fig:cartoonIB}Schematic evolution of an isolated binary which can give birth to a merging BH (see e.g. \cite{belczynski2016,mapellietal2017,stevenson2017,giacobbo2018}). }
\end{figure}
%%%%%%%%%%%%%%%%%%%%%%%%%%%%%%%%%%%%%%%%%%%%%%%%%%%%%%%%%%%%%%%%%%%%%%%%%%%%

%To summarize, we would like to illustrate schematically what is the most popular ``isolated binary'' evolutionary scenario for merging BHs like GW150914 and the other massive BHs (see e.g. \cite{belczynski2016,stevenson2017,mapelli2017,giacobbo2018} observed by the LIGO-Virgo collaboration \cite{abbott2016a,abbott2016c,abbott2017a,abbott2017b}. Figure~\ref{fig:cartoonIB}
To summarize, let us illustrate schematically the evolution of an isolated stellar binary (see e.g. \cite{belczynski2016,mapellietal2017,stevenson2017,giacobbo2018}) which can give birth to merging BHs like GW150914 and the other massive BHs  observed by the LIGO-Virgo collaboration \cite{abbott2016a,abbott2016c,abbott2017a,abbott2017b}. In the following discussion, several details of stellar evolution have been simplified or skipped to facilitate the reading for non specialists.

Figure~\ref{fig:cartoonIB} shows the evolution of an isolated binary system composed of two massive stars. These stars are gravitationally bound since their birth. %, i.e. they formed from the same molecular cloud.
Initially, the two stars are both on the main sequence (MS). When the most massive one leaves the MS (i.e. when Hydrogen burning in the core is over, which happens usually on a time-scale of few Myr for massive stars with ZAMS mass $m_{\rm ZAMS}\gtrsim{}30$ M$_\odot$), its radius starts inflating and can grow by a factor of a hundreds. %If the orbital separation is smaller or comparable to this value, the two stars merge leaving a single star.
The most massive star becomes a giant star with a Helium core and a large Hydrogen envelope. If its radius equals the Roche lobe (equation~\ref{eq:rlobe}), the system starts a stable mass-transfer episode. Some mass is lost by the system, some is transferred to the companion. After several additional evolutionary stages, the primary collapses to a BH (a direct collapse is preferred with respect to a SN explosion if we want the BH to be rather massive). At this stage the system is still quite large (hundreds to thousands of solar radii).  

When also the secondary leaves the MS, growing in radius, the system enters a CE phase: the BH and the Helium core spiral in. If the orbital energy is not sufficient to unbind the envelope, then the BH merges with the Helium core leaving a single BH. In contrast, if the envelope is ejected, we are left with a new binary, composed of the BH and of a stripped naked Helium star. The new binary has a much smaller orbital separation (tens of solar radii) than the pre-CE binary, because of the spiral-in. If this new binary remains bound after the naked Helium star undergoes a SN explosion or if the naked Helium star is sufficiently massive to directly collapse to a BH, the system evolves into a close BHB, which might merge within a Hubble time.

The most critical quantities in this scenario are the masses of the two stars and also their initial separation (with respect to the stellar radii): a BHB can merge within a Hubble time only if its initial orbital separation is tremendously small (few tens of solar radii, unless the eccentricity is rather extreme); but a massive star ($>20$ M$_\odot$) can reach a radius of several thousand solar radii during its evolution. Thus, if the initial orbital separation of the stellar binary is tens of solar radii, the binary merges before it can become a BHB. On the other hand, if the initial orbital separation is too large, the two BHs will never merge. In this scenario, the two BHs can merge only if the initial orbital separation of the progenitor stellar binary is in the range which allows the binary to enter CE and then to leave a short period BHB. This range of initial orbital separations dramatically depends on CE efficiency and on the details of stellar mass and radius evolution.

\section{The dynamics of black hole binaries}
In the previous sections of this review, we discussed the formation of BH binaries as isolated binaries. There is an alternative channel for BH binary formation: the dynamical evolution scenario.

\subsection{Dynamically active environments}
Collisional dynamics is important for the evolution of binaries only if they are in a dense environment ($\gtrsim{}10^3$ stars pc$^{-3}$), such as a star cluster. On the other hand, astrophysicists believe that the vast majority of massive stars (which are BH progenitors) form in star clusters \cite{lada2003,weidner2006,weidner2010,portegieszwart2010}. 

Most studies of dynamical formation of BH binaries focus on globular clusters (e.g. \cite{sigurdsson1993,portegieszwart2000,mapelli2005,downing2010,downing2011,benacquista2013,tanikawa2013,samsing2014,rodriguez2015,rodriguez2016,askar2017,samsing2017,samsing2018,askar2018,rodriguez2018}). {\it Globular clusters} are old stellar systems ($\sim{}12$ Gyr), mostly very massive ($>10^4$ M$_\odot$) and dense ($>10^4$ M$_\odot$ pc$^{-3}$). They are sites of intense dynamical processes (such as the gravothermal catastrophe). However, globular clusters represent a tiny fraction of the baryonic mass in the local Universe ($\lesssim{}1$ per cent, \cite{harris2013}).

In contrast, only few studies of BH binaries (e.g \cite{ziosi2014,mapelli2016,kimpson2016,banerjee2017a,banerjee2017b,fujii2018}) focus on {\it young star clusters}. These young ($\lesssim{}100$ Myr), relatively dense ($>10^3$ M$_\odot$ pc$^{-3}$) stellar systems are thought to be the most common birthplace of massive stars. When they evaporate (by gas loss) or are disrupted by the tidal field of their host galaxy, their stellar content is released into the field. Thus, it is reasonable to expect that a large fraction of BH binaries which are now in the field may have formed in young star clusters, where they participated in the dynamics of the cluster. The reason why young star clusters have been neglected in the past is exquisitely numerical: the dynamics of young star clusters needs to be studied with direct N-body simulations, which are rather expensive (they scale as $N^2$), combined with population-synthesis simulations. Moreover, their dynamical evolution may be significantly affected by the presence of gas. Including gas would require a challenging interface between direct N-body simulations and hydrodynamical simulations, which has been done in very few cases \cite{moeckel2010,fujii2015,fujii2016,parker2013,parker2015a,parker2015b,parker2017,mapelli2017} and has been never used to study BH binaries. Finally, a fraction of young star clusters might survive gas evaporation and tidal disruption and evolve into older {\it open clusters}, like M67.

Another flavour of star cluster where BH binaries might form and evolve dynamically are {\it nuclear star clusters}, i.e. star clusters which lie in the nuclei of galaxies. Nuclear star clusters are rather common in galaxies (e.g. \cite{boeker2002,ferrarese2006,graham2009}), are usually more massive and denser than globular clusters, and may co-exist with super-massive BHs (SMBHs). Stellar-mass BHs formed in the innermost regions of a galaxy could even be ``trapped'' in the accretion disc of the central SMBH, triggering their merger (see e.g. \cite{stone2016,bartos2017,mckernan2017}). These features make nuclear star clusters unique among star clusters, for the effects that we will describe in the next sections.

%%%%%%%%%%%%%%%%%%%%%%%%%%%%%%%%%%%%FIGURE%%%%%%%%%%%%%%%%%%%%%%%%%%%%%%%%%%%%
%{\bf fare una figura collage di diversi star clusters}
%%%%%%%%%%%%%%%%%%%%%%%%%%%%%%%%%%%%%%%%%%%%%%%%%%%%%%%%%%%%%%%%%%%%%%%%%%%%%%

\subsection{Three-body encounters}
We now review what are the main dynamical effects which can affect a BH binary, starting from three-body encounters.

Binaries have a energy reservoir, their internal energy:
\begin{equation}
E_{\rm int}=\frac{1}{2}\,{}\mu{}\,{}v^2-\frac{G\,{}m_1\,{}m_2}{r},
\end{equation}
where $\mu{}=m_1\,{}m_2/(m_1+m_2)$ is the reduced mass of the binary (whose components have mass $m_1$ and $m_2$), $v$ is the relative velocity between the two members of the binary, and $r$ is the distance between the two members of the binary. As shown by Kepler's laws \cite{kepler1621}, $E_{\rm int}=-E_{\rm b}=-G\,{}m_1\,{}m_2/(2\,{}a)$, where $E_{\rm b}$ is the binding energy of the binary ($a$ being the semi-major axis of the binary).

The internal energy of a binary can be exchanged with other stars only if the binary undergoes a close encounter with a star, so that its orbital parameters are perturbed by the intruder. This happens only if a single star approaches the binary by few times its orbital separation. We define this close encounter between a binary and a single star as a {\it three-body encounter}. For this to happen with a non-negligible frequency, the binary must be in a dense environment, because the rate of three-body encounters scales with the local density of stars.

Three-body encounters have crucial effects on BH binaries, such as {\it exchanges}, {\it hardening}, and {\it ejections}. 

\subsection{Exchanges}
\begin{figure}
\center{
\includegraphics[width=12cm]{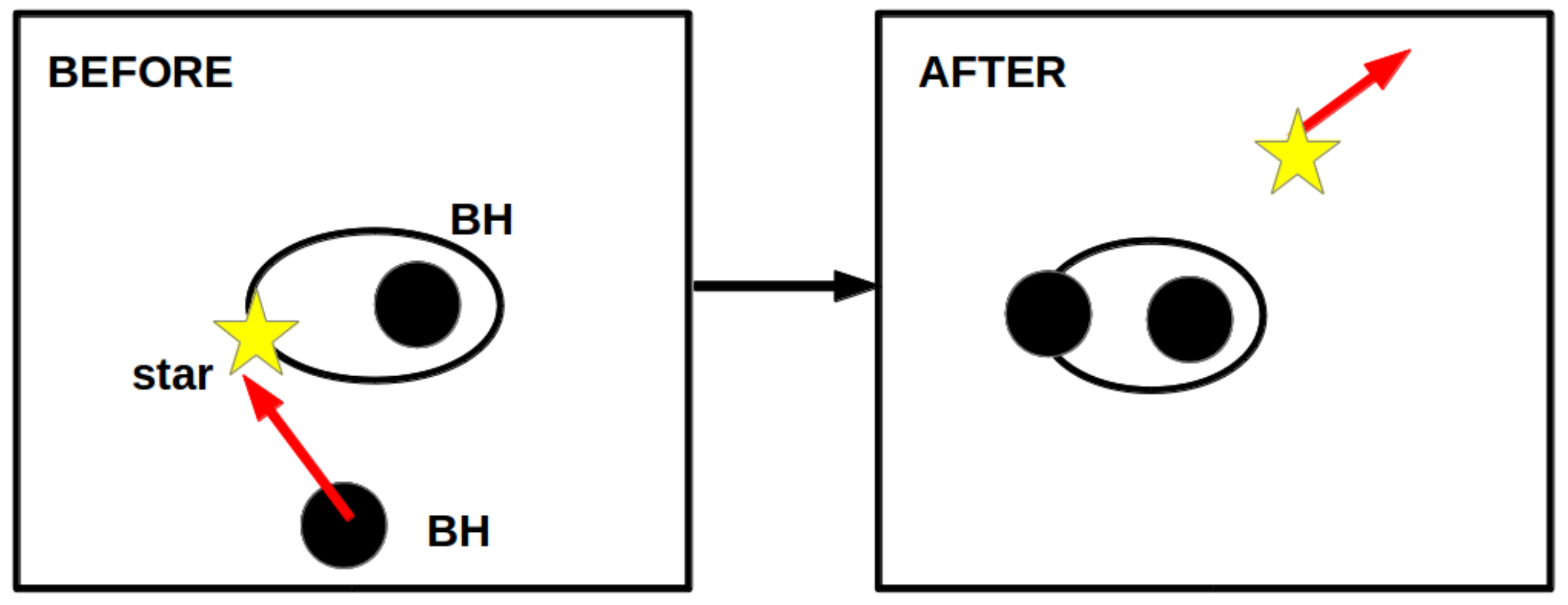}     % includes figure foo.eps
}
\caption{\label{fig:exchange}Cartoon of a dynamical exchange. A binary composed of a BH and a star interacts with a BH, which replaces the star.}
\end{figure}

Dynamical exchanges are three-body encounters during which one of the former members of the binary is replaced by the intruder (see Figure~\ref{fig:exchange}).
%%%%%%%%%%%%%%%%%%%%%%%%%%%%%FIGURE %%%%%%%%%%%%%%%%%%%%%%%%%%%%%%%%%%%%%%%%
%%%%%%%%%%%%%%%%%%%%%%%%%%%%%%%%%%%%%%%%%%%%%%%%%%%%%%%%%%%%%%%%%%%%%%%%%%%%

Exchanges may lead to the formation of new double BH binaries. As it is shown in Figure~\ref{fig:exchange}, if a binary composed of a BH and a low-mass star undergoes an exchange with a single BH, this leads to the formation of a new double BH binary. This is a very important difference between BHs in the field and in star clusters: a BH which forms as a single object in the field has negligible chances to become member of a binary system, while a single BH in the core of a star cluster has good chances of becoming member of a binary by exchanges.

Exchanges are expected to lead to the formation of many more double BH binaries than they can destroy, because the probability for an intruder to replace one of the members of a binary is $\approx{}0$ if the intruder is less massive than both binary members, while it suddenly jumps to $\sim{}1$ if the intruder is more massive than one of the members of the binary \cite{hills1980}. Since BHs are among the most massive bodies in a star cluster (after their massive progenitors transform into them), they are very efficient in acquiring companions through dynamical exchanges. 

Thus, exchanges are a crucial mechanism to form BH binaries dynamically. By means of direct N-body simulations, Ziosi et al. (2014, \cite{ziosi2014}) show that $>90$ per cent of double BH binaries in young star clusters form by dynamical exchange.

Moreover, BH binaries formed via dynamical exchange will have some distinctive features with respect to field BH binaries (see e.g. \cite{ziosi2014}):
\begin{itemize}
\item{}double BH binaries formed by exchanges will be (on average) more massive than isolated double BH binaries, because more massive intruders have higher chances to acquire companions;
\item{}exchanges trigger the formation of highly eccentric double BH binaries (eccentricity is then significantly reduced by circularisation induced by GW emission, if the binary enters the regime where GW emission is effective);
\item{}double BH binaries born by exchange will likely have misaligned spins, because exchanges tend to randomize the spins.
\end{itemize}

%For example, Figure~\ref{fig:rodriguez} shows the distribution of initial eccentricities of BH binaries in globular clusters versus the eccentricity they have when they emit GWs at $10$ Hz (i.e. when they enter LIGO--Virgo's band). While the initial eccentricity is close to $e\sim{}1$ for most systems, the eccentricity in the LIGO's band is very close to zero.
%%%%%%%%%%%%%%%%%%%%%%%%%%%%%FIGURE %%%%%%%%%%%%%%%%%%%%%%%%%%%%%%%%%%%%%%%%
%\begin{figure}
%\center{
%\includegraphics[width=6cm]{rodriguez}     % includes figure foo.eps
%}
%\caption{\label{fig:rodriguez}Top panel: initial eccentricity of BH binaries versus eccentricity when the binary emits GWs at 10 Hz (i.e. when it enters LIGO-Virgo's frequency range), corresponding to an orbital frequency of 5 Hz. Bottom panel: probability distribution of eccentricities when the binary emits GWs at 10 Hz. From Figure~10 of Rodriguez et al. (2016, \cite{rodriguez2016}).}
%\end{figure}
%%%%%%%%%%%%%%%%%%%%%%%%%%%%%%%%%%%%%%%%%%%%%%%%%%%%%%%%%%%%%%%%%%%%%%%%%%%%

Zevin et al. (2017, \cite{zevin2017}) compare a set of simulations of field binaries with a set of simulations of globular cluster binaries, run with the same population-synthesis code. The most striking difference between merging BH binaries in their globular cluster simulations and in their population-synthesis simulations is the dearth of merging BHs with mass $<10$ M$_\odot$ in the globular cluster simulations. This is likely due to the fact that exchanges tend to destroy binaries composed of light BHs.

Spin misalignments are another possible feature to discriminate between field binaries and star cluster binaries (e.g. \cite{farr2017a,farr2017b}). Unfortunately, there is no robust theory to predict the magnitude of the spin of a BH given the spin of its parent star \cite{millermiller2015}. In principle, we do not know whether BH spins are low ($\sim{}0$) or high ($\sim{}1$), given the spin of the parent star. However, we can reasonably state that the orientation of the spin of a BH matches the orientation of the spin of its progenitor star, if the latter evolved in isolation and directly collapsed to a BH.

Thus, we expect that an isolated binary in which the secondary becomes a BH by direct collapse results in a double BH binary with aligned spins (i.e. the spins of the two BHs have the same orientation, which is approximately the same as the orbital angular momentum direction of the binary), because tidal evolution and mass transfer in a binary tend to synchronise the spins \cite{hurley2002}. On the other hand, if the secondary undergoes a SN explosion, the natal kick may reshuffle spins.

For dynamically formed BH binaries (through exchange) we expect misaligned, or even nearly isotropic spins, because any original spin alignment is completely reset by three-body encounters.

Currently, we have only poor constraints on BH spins for the first GW detections. LIGO-Virgo have provided a measurement for the effective spin $\chi_{\rm eff}$ which is the sum of the components of the spins of the two BHs along the orbital angular momentum vector of the binary:
\begin{equation}\label{eq:effectivespin}
\chi_{\rm eff}=\frac{c}{G\,{}(m_1+m_2)}\,{}\left(\frac{\vec{S}_1}{m_1}+\frac{\vec{S}_2}{m_2}\right)\dot{}\frac{\vec{L}}{|\vec{L}|}\equiv{}\frac{1}{m_1+m_2}\,{}(m_1\,{}\chi_1+m_2\,{}\chi_2),
\end{equation}
where $\vec{S}_1$ and $\vec{S}_2$ are the spin angular momentum vectors of the BHs, $\vec{L}$ is the orbital angular momentum of the binary, $\chi{}_1$ and $\chi{}_2$ are the dimensionless projections of the individual BH spins, respectively. By construction, $-1\leq{}\chi_{\rm eff}\leq{}1$. A positive (negative) value of $\chi_{\rm eff}$ means that the spins of the two BHs are aligned (counter-aligned).

Only for GW151226, the measured value of $\chi_{\rm eff}$ is significantly larger than zero \cite{abbott2016b}, indicating (at least partial) alignment. For GW150914 \cite{abbott2016a}, LVT151012 \cite{abbott2016c}, GW170104 \cite{abbott2017a}, GW170608 \cite{abbott2017c} and GW170814 \cite{abbott2017b}, the effective spin is consistent with zero and may be either negative or positive. With a Bayesian approach, \cite{farr2017a} estimate that an isotropic distribution of spins can be preferred with respect to an aligned distribution of spins at 2.4 $\sigma{}$ confidence level, based on the first three detections and on the candidate event LVT151012.

\subsection{hardening}
If a double BH binary undergoes a number of three-body encounters during its life, we expect that its semi-major axis will shrink as an effect of the encounters. This process is called dynamical {\it hardening}.

Following \cite{heggie1975}, we call hard binaries (soft binaries) those binaries with binding energy larger (smaller) than the average kinetic energy of a star in the star cluster. According to Heggie's law \cite{heggie1975}, hard binaries tend to harden (i.e. to become more and more bound) via three-body encounters. In other words, a fraction of the internal energy of a hard binary can be transferred into kinetic energy of the intruders and of the centre-of-mass of the binary during three body encounters. This means that the binary loses internal energy and its semi-major axis shrinks. 

Most double BH binaries are expected to be hard binaries, because BHs are among the most massive bodies in star clusters. Thus, double BH binaries are expected to harden as a consequence of three-body encounters. The hardening process may be sufficiently effective to shrink a BH binary till it enters the regime where GW emission is efficient: a BH binary which is initially too loose to merge may then become a GW source thanks to dynamical hardening.

%%%%%%%%%%%%%%%%%%%%%%%%%%%%%FIGURE %%%%%%%%%%%%%%%%%%%%%%%%%%%%%%%%%%%%%%%%
\begin{figure}
\center{
\includegraphics[width=13cm]{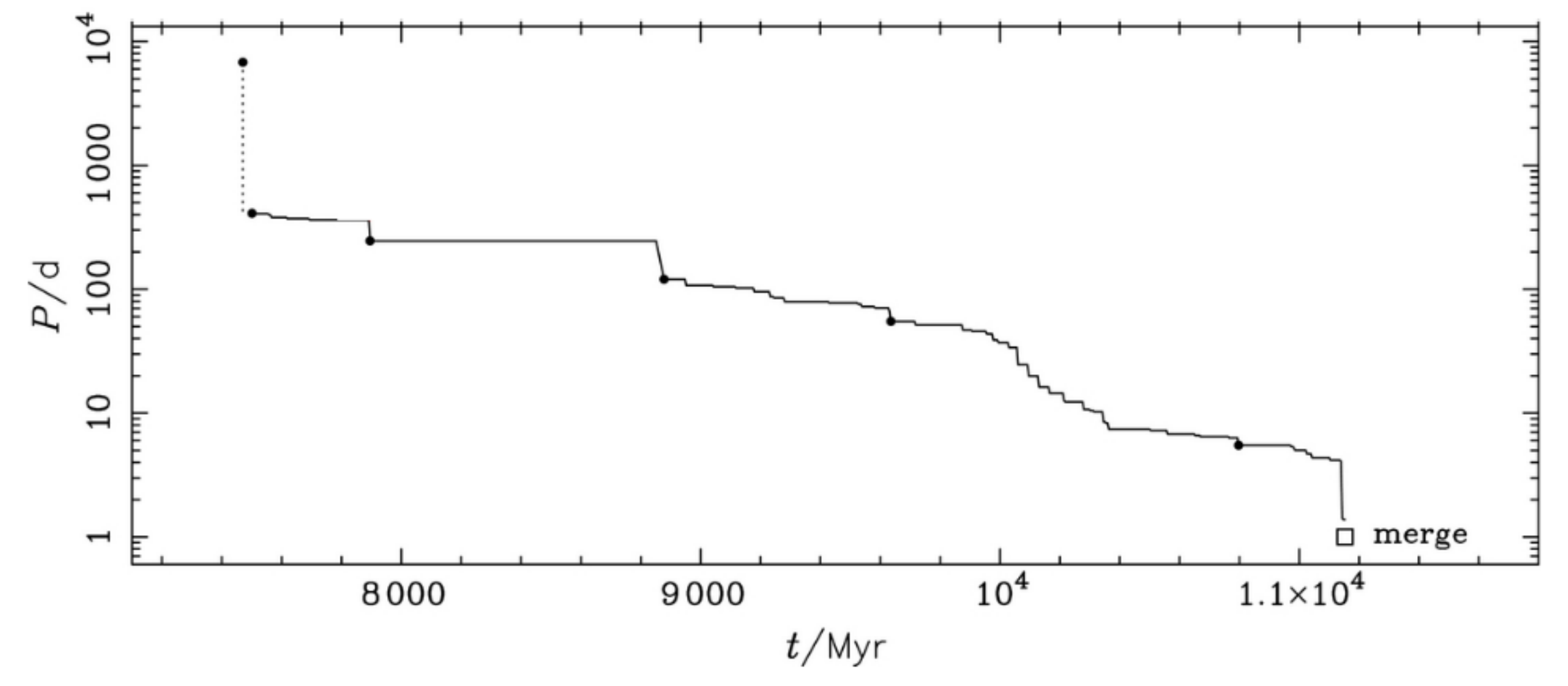}     % includes figure foo.eps
}
\caption{\label{fig:hurley}Time evolution of the orbital period of a simulated BH binary (from an N-body simulation of a globular cluster). Adapted from Figure 3 of Hurley et al. 2016 \cite{hurley2016}.}
\end{figure}
%%%%%%%%%%%%%%%%%%%%%%%%%%%%%%%%%%%%%%%%%%%%%%%%%%%%%%%%%%%%%%%%%%%%%%%%%%%%
Figure~\ref{fig:hurley} is a clear example of this process. It shows the period evolution of a simulated double BH binary in a globular cluster. The orbital period decreases till the binary eventually merges. The decrease of the orbital period is not smooth, but proceeds by smaller and larger steps, which indicate hardening due to three-body encounters and even few exchanges.

It is even possible to make a simple analytic estimate of the evolution of the semi-major axis of a double BH binary which is affected by three-body encounters and by GW emission (equation~9 of \cite{colpi2003}):
\begin{equation}\label{eq:miatesi}
\frac{da}{dt}=-2\,{}\pi{}\,{}\xi{}\,{}\frac{G\,{}\rho{}}{\sigma{}}\,{}a^2-\frac{64}{5}\frac{G^3\,{}m_1\,{}m_2\,{}(m_1+m_2)}{c^5\,{}(1-e^2)^{7/2}}\,{}a^{-3},
\end{equation}
where $\xi{}\sim{}0.2-1$ is a dimensionless hardening parameter (which has been estimated through numerical experiments, \cite{hills1983}), $\rho{}$ is the local mass density of stars, $\sigma{}$ is the local velocity dispersion, $c$ is the light speed, $e$ is the eccentricity. The first part of the right-hand term of equation~\ref{eq:miatesi} accounts for the effect of three-body hardening on the semi-major axis. It scales as $da/dt\propto{}-a^2$, indicating that the larger the binary is, the more effective the hardening. This can be easily understood considering that the geometric cross section for three body interactions with a binary scales as $a^2$.

The second part of the right-hand term of equation~\ref{eq:miatesi} accounts for energy loss by GW emission. It is the first-order approximation of the calculation by Peters (1964, \cite{peters1964}). It scales as $da/dt\propto{}-a^{-3}$ indicating that GW emission becomes efficient only when the two BHs are very close to each other.

%%%%%%%%%%%%%%%%%%%%%%%%%%%%%FIGURE %%%%%%%%%%%%%%%%%%%%%%%%%%%%%%%%%%%%%%%%
\begin{figure}
\center{
\includegraphics[width=12cm]{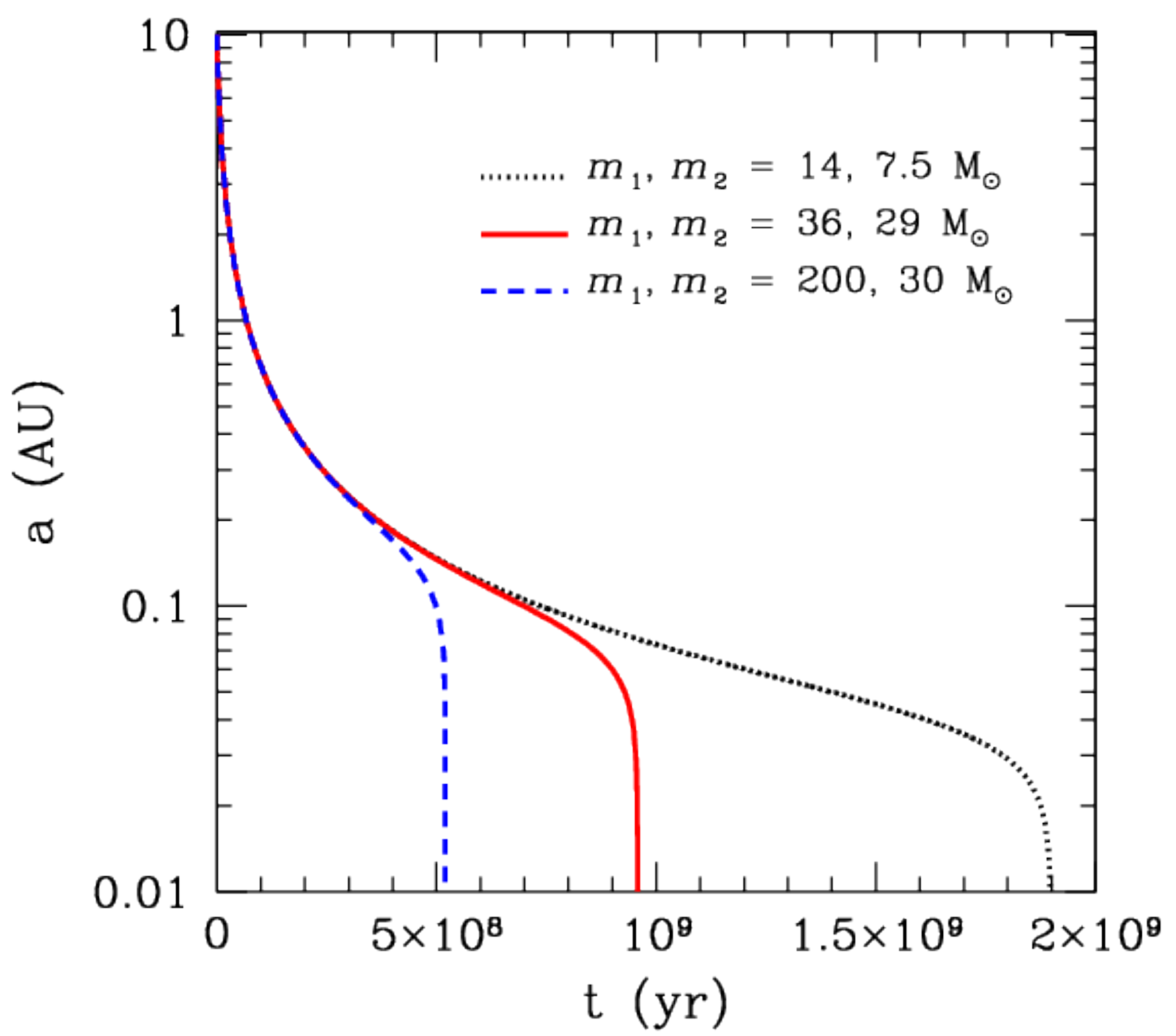}     % includes figure foo.eps
}
\caption{\label{fig:miatesi}Time evolution of the semi-major axis of three BH binaries estimated from equation~\ref{eq:miatesi}. Blue dashed line: BH binary with masses $m_1=200$ M$_\odot$, $m_2=30$ M$_\odot$; red solid line: $m_1=36$ M$_\odot$, $m_2=29$ M$_\odot$; black dotted line: $m_1=14$ M$_\odot$, $m_2=7.5$ M$_\odot$. For all BH binaries: $\xi{}=1$, $\rho{}=10^5$ M$_\odot$ pc$^{-3}$, $\sigma{}=10$ km s$^{-1}$, $e=0$ (here we assume that $\rho{}$, $\sigma{}$ and $e$ do not change during the evolution), initial semi-major axis of the BH binary $a_{\rm i}=10$ AU.}
\end{figure}
%%%%%%%%%%%%%%%%%%%%%%%%%%%%%%%%%%%%%%%%%%%%%%%%%%%%%%%%%%%%%%%%%%%%%%%%%%%%

In Figure~\ref{fig:miatesi} we solve equation~\ref{eq:miatesi} numerically for three double BH binaries with different mass. All binaries evolve through (i) a first phase in which hardening by three body encounters dominates the evolution of the binary, (ii) a second phase in which the semi-major axis stalls because three-body encounters become less efficient as the semi-major axis shrink, but the binary is still too large for GW emission to become efficient, and (iii) a third phase in which the semi-major axis drops because the binary enters the regime where GW emission is efficient.

\subsection{Dynamical ejections}
During three-body encounters, a fraction of the internal energy of a hard binary is transferred into kinetic energy of the intruders and of the centre-of-mass of the binary. As a consequence, the binary and the intruder recoil. The recoil velocity is generally of the order of few km s$^{-1}$, but can be up to several hundred km s$^{-1}$. 

Since the escape velocity from a globular cluster is $\sim{}30$ km s$^{-1}$ and the escape velocity from a young star cluster or an open cluster is even lower, both the recoiling binary and the intruder can be ejected from the parent star cluster. If the binary and/or the intruder are ejected, they become field objects and cannot participate in the dynamics of the star cluster anymore. Thus, not only the ejected double BH binary stops hardening, but also the ejected intruder, if it is another compact object, loses any chance of entering a new binary by dynamical exchange.

\begin{figure}
\center{
\includegraphics[width=12cm]{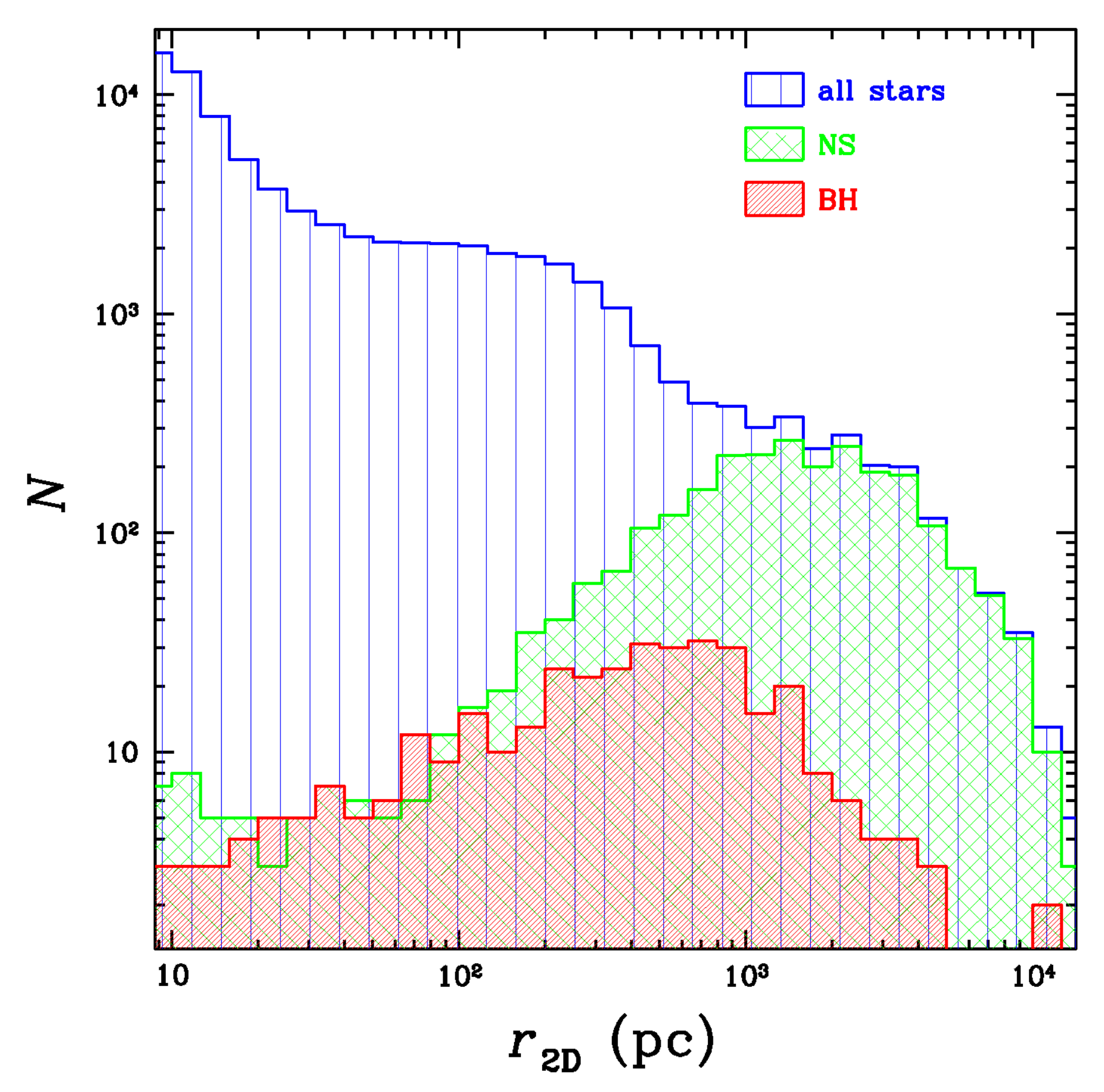}     % includes figure foo.eps
}
\caption{\label{fig:ejection}Distribution of 2 dimensional distances ($r_{\rm 2D}$) of stars from the centre of the parent star cluster 100 Myr after the beginning of the simulation. This plot combines 300 simulations of young star clusters discussed in \cite{mapelli2013}. All stars shown in the Figure have distance $>10$ pc from the centre of the star cluster, indicating that they were ejected. Blue histogram: all ejected stars; green histogram: ejected NSs; red histogram: ejected BHs. NSs are ejected both by SN kick ($\sim{}50$ per cent) and by dynamical recoil ($\sim{}50$ per cent), while most BHs are ejected by dynamical recoil.}
\end{figure}

Figure~\ref{fig:ejection} shows that this process is particularly efficient in young star clusters, especially for NSs (which are lighter than BHs). In 100 Myr, about 90 per cent of NSs and 40 per cent of BHs are ejected from the parent cluster, either by dynamical recoil or by SN kick (\cite{mapelli2013}, see also \cite{mapelli2011,downing2011,pavlik2018}). 

Dynamical ejections of double NSs and of BH-NS binaries were proposed to be one of the possible explanations for the host-less short gamma-ray bursts, i.e. gamma-ray bursts whose position in the sky appears to be outside any observed galaxy \cite{fong2013}. Host-less bursts may be $\sim{}25$ per cent of all short gamma-ray bursts.
%%%%%%%%%%%%%%%%%%%%%%%%%%%%%FIGURE %%%%%%%%%%%%%%%%%%%%%%%%%%%%%%%%%%%%%%%%
%%%%%%%%%%%%%%%%%%%%%%%%%%%%%%%%%%%%%%%%%%%%%%%%%%%%%%%%%%%%%%%%%%%%%%%%%%%%

In general, ejections of compact objects and compact-object binaries from their parent star cluster can be the result of at least three different processes:
\begin{itemize}
\item{}dynamical ejections;
\item{}SN kicks \cite{hobbs2005,fryer2012};
\item{}GW recoil \cite{lousto2009,campanelli2007,gonzalez2007}.
\end{itemize}

GW recoil is a relativistic kick occurring when a BH binary merges. It results in kick velocities up to thousands of km s$^{-1}$ and usually of the order of hundreds of km s$^{-1}$.

Ejections (by dynamics, SN kick or GW recoil) may be the main process at work against mergers of second-generation BHs, where for second-generation BHs we mean BHs which were born from the merger of two BHs rather than from the collapse of a star \cite{gerosaberti2017}. In globular clusters, open clusters and young star clusters, a BH binary has good chances of being ejected by three-body encounters before it merges (see \cite{millerhamilton2002} for a detailed calculation) and a very high chance of being ejected by GW recoil after it merges. The only place where merging BHs can easily avoid ejection by GW recoil are nuclear star clusters, whose escape velocity can reach hundreds of km s$^{-1}$.

%\subsection{Spitzer's instability}

\subsection{Formation of intermediate-mass black holes by runaway collisions}
In Section~\ref{sec:remnants}, we have mentioned that intermediate-mass black holes (IMBHs, i.e. BHs with mass $100\lesssim{}m_{\rm BH}\lesssim{}10^4$ M$_\odot$) might form from the direct collapse of metal-poor extremely massive stars \cite{spera2017}. Other formation channels have been proposed for IMBHs and most of them involve dynamics of star clusters. The formation of massive BHs by runaway collisions has been originally proposed about half a century ago \cite{colgate1967,sanders1970} and was then elaborated by several authors (e.g. \cite{portegieszwart1999,portegieszwart2002,portegieszwart2004,gurkan2006,freitag2006,mapelli2006,mapelli2008,giersz2015,mapelli2016}). 

%%%%%%%%%%%%%%%%%%%%%%%%%%%%%FIGURE %%%%%%%%%%%%%%%%%%%%%%%%%%%%%%%%%%%%%%%%
\begin{figure}
\center{
\includegraphics[width=14cm]{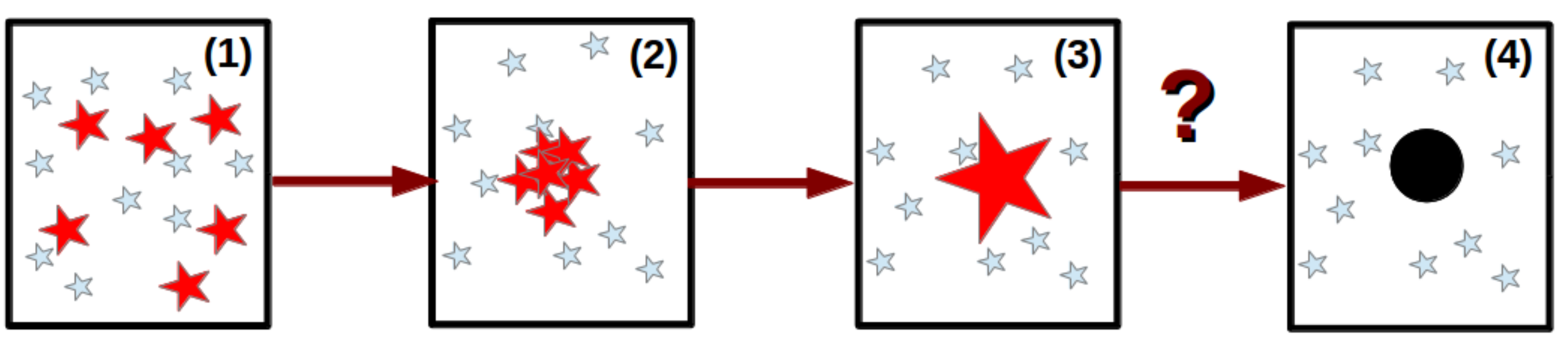}     % includes figure foo.eps
}
\caption{\label{fig:runaway}Cartoon of the runaway collision scenario in dense young star clusters (see e.g. \cite{portegieszwart2002}). From left to right: (1) the massive stars (red big stars) and the low-mass stars (blue small stars) follow the same initial spatial distribution; (2) dynamical friction leads the massive stars to sink to the core of the cluster, where they start colliding between each other; (3) a very massive star ($\gg{}100$ M$_\odot$) forms as a consequence of the runaway collisions; (4) this massive star might be able to directly collapse into a BH.}
\end{figure}
%%%%%%%%%%%%%%%%%%%%%%%%%%%%%%%%%%%%%%%%%%%%%%%%%%%%%%%%%%%%%%%%%%%%%%%%%%%%

The basic idea is the following (as summarized by the cartoon in Figure~\ref{fig:runaway}). In a dense star cluster, dynamical friction \cite{chandrasekhar1943} makes massive stars to decelerate because of the drag exerted by lighter bodies, on a timescale which can be expressed as
\begin{equation}
t_{\rm DF}(M)=\frac{\langle{}m\rangle{}}{M}\,{}t_{\rm rlx},
\end{equation}
where $\langle{}m\rangle{}$ is the average star mass in a star cluster (for young star clusters $\langle{}m\rangle{}\sim{}1$ M$_\odot$), $M$ is the mass of the massive star we consider ($M>\langle{}m\rangle{}$) and $t_{\rm rlx}$ is the central two body relaxation timescale of the star cluster (i.e. the timescale needed for a star to completely lose memory of its initial velocity as an effect of two-body encounters, \cite{spitzer1971}). For dense young star clusters \cite{portegieszwart2010}
\begin{equation}
t_{\rm rlx}\simeq{}20\,{}{\rm Myr}\,{}\left(\frac{M_{\rm cl}}{10^4{\rm M}_\odot}\right)^{1/2}\,{}\left(\frac{R_{\rm cl}}{1\,{}{\rm pc}}\right)^{3/2}\,{}\left(\frac{{\rm M}_\odot}{\langle{}m\rangle{}}\right),
\end{equation}
where $M_{\rm cl}$ is the total star cluster mass and $R_{\rm cl}$ is the virial radius of the star cluster.

Thus, for a star with mass $M\ge{}25$ M$_\odot$, we estimate $t_{\rm DF}\le{}1$ Myr: dynamical friction is very effective in dense massive young star clusters. Because of dynamical friction, massive stars segregate to the core of the cluster, which is the centre of the cluster potential well.

If the most massive stars in a dense young star cluster sink to the centre of the cluster by dynamical friction on a time shorter than their lifetime (i.e. before core-collapse SNe take place, removing a large fraction of their mass), then the density of massive stars in the cluster core becomes extremely high. This makes collisions between massive stars extremely likely. Actually, direct N-body simulations show that collisions between massive stars proceed in a runaway sense, leading to the formation of a very massive ($\gg{}100$ M$_\odot$) star \cite{portegieszwart2002}. The main open question is: ``What is the final mass of the collision product? Is the collision product going to collapse to an IMBH?''.

%The basic idea (see Figure~\ref{fig:runaway} for a cartoon) is that if the most massive stars in a dense young star cluster sink to the centre of the cluster by dynamical friction on a time shorter than their lifetime (i.e. before core-collapse SNe take place, removing a large fraction of their mass), then the density of massive stars in the cluster core becomes extremely high. This makes collisions between massive stars extremely likely. Actually, direct N-body simulations show that collisions between massive stars proceed in a runaway sense, leading to the formation of a very massive ($\gg{}100$ M$_\odot$) star \cite{portegieszwart2002}. The main open question is: ``What is the final mass of the collision product? Is the collision product going to collapse to an IMBH?''.

There are essentially two critical issues: (i) how much mass is lost during the collisions? (ii) how much mass does the very-massive star lose by stellar winds?  

Hydrodynamical simulations of colliding stars \cite{gaburov2008,gaburov2010} show that massive star can lose $\approx{}25$ per cent of their mass during collisions. Even if we optimistically assume that no mass is lost during and immediately after the collision (when the collision product relaxes to a new equilibrium), the resulting very massive star will be strongly radiation pressure dominated and is expected to lose a significant fraction of its mass by stellar winds. Recent studies including the effect of the Eddington factor on mass loss \cite{mapelli2016,spera2017} show that IMBHs cannot form from runaway collisions at solar metallicity. At lower metallicity ($Z\lesssim{}0.1$ Z$_\odot$) approximately $10-30$ per cent of runaway collision products in young dense star clusters can become IMBHs by direct collapse (they also avoid being disrupted by pair-instability SNe).

The majority of runaway collision products do not become IMBHs but they end up as relatively massive BHs ($\sim{}20-90$ M$_\odot$, \cite{mapelli2016}). If they remain inside their parent star cluster, such massive BHs are extremely efficient in acquiring companions by dynamical exchanges. Mapelli (2016, \cite{mapelli2016}) find that all stable binaries formed by the runaway collision product are double compact object binaries and thus are possibly important sources of GWs in the LIGO-Virgo range.

\subsection{Formation of intermediate-mass black holes by repeated mergers}
The runaway collision scenario occurs only in the early stages of the evolution of a star cluster, when massive stars are still alive. However, it has been proposed that IMBHs form even in old clusters (e.g. globular clusters) by repeated mergers of smaller BHs (e.g. \cite{millerhamilton2002,giersz2015}).

%%%%%%%%%%%%%%%%%%%%%%%%%%%%%FIGURE %%%%%%%%%%%%%%%%%%%%%%%%%%%%%%%%%%%%%%%%
\begin{figure}
\center{
\includegraphics[width=14cm]{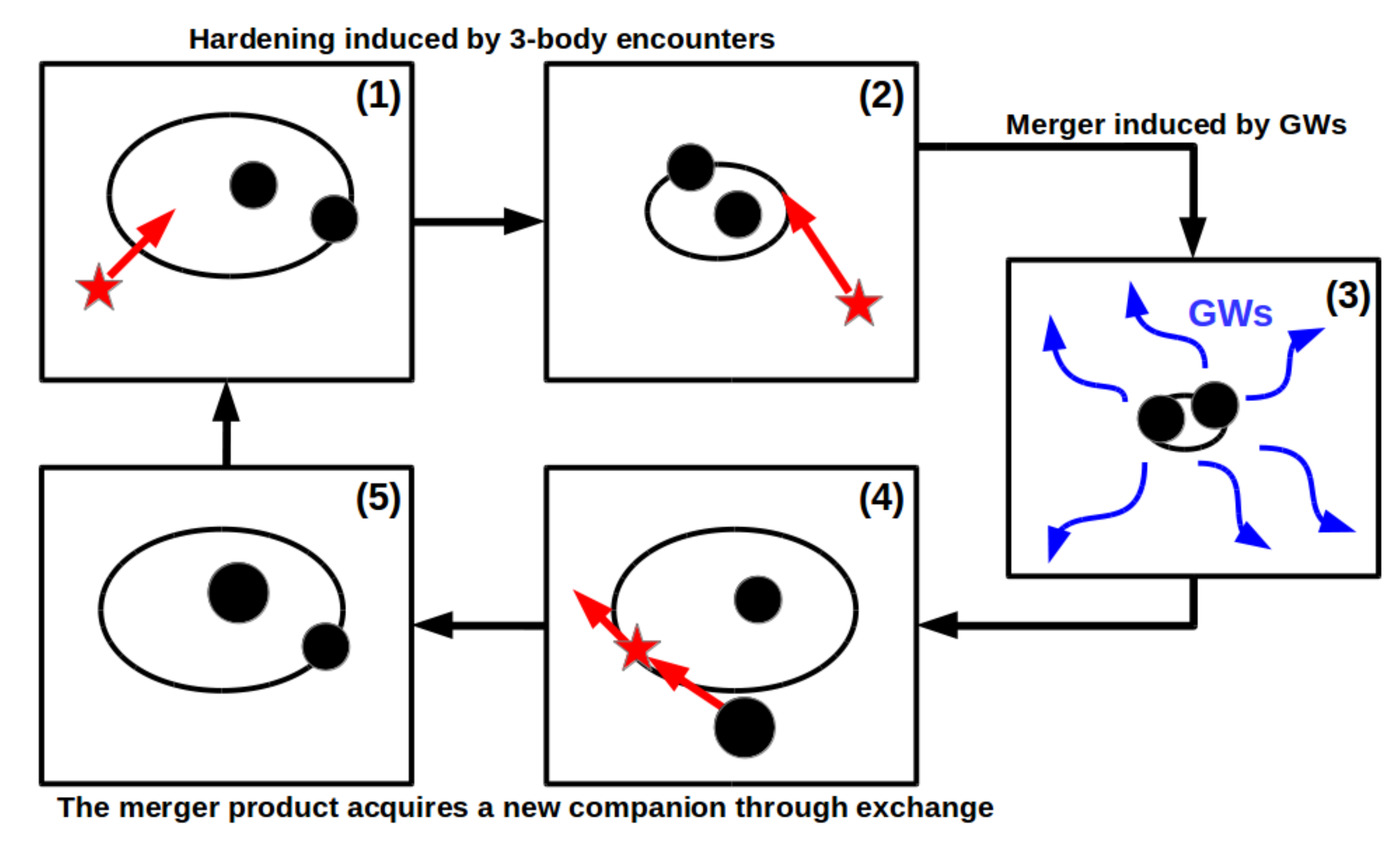}     % includes figure foo.eps
}
\caption{\label{fig:repeatedmergers}Cartoon of the repeated merger scenario in old star clusters (see e.g. \cite{millerhamilton2002,giersz2015}). From top to bottom and from left to right: (1) a BH binary undergoes three-body encounters in a star cluster; (2) three-body encounters harden the BH binary, shrinking its semi-major axis; (3) the BH binary hardens by three-body encounters till it enters the regime where GW emission is efficient: the binary semi-major axis decays by GW emission and the binary merges; (4) a single bigger BH forms as result of the merger, which may acquire a new companion by dynamical exchange (if it is not ejected by GW recoil); (5) the new binary containing the bigger BH starts shrinking again by three body encounters (1). This loop may be repeated several times till the main BH becomes an IMBH.
}
\end{figure}
%%%%%%%%%%%%%%%%%%%%%%%%%%%%%%%%%%%%%%%%%%%%%%%%%%%%%%%%%%%%%%%%%%%%%%%%%%%%

The simple idea is illustrated in Figure~\ref{fig:repeatedmergers}. A stellar BH binary in a star cluster is usually a rather hard binary. Thus, it shrinks by dynamical hardening till it may enter the regime where GW emission is effective. In this case, the BH binary merges leaving a single more massive BH. Given its relatively large mass, the new BH has good chances to acquire a new companion by exchange. Then, the new BH binary starts hardening again by three-body encounters and the story may repeat several times, till the main BH becomes an IMBH.

This scenario has one big advantage: it does not depend on stellar evolution, so we are confident that the BH will grow in mass by mergers, if it remains inside the cluster. However, there are several issues. First, the BH binary may be ejected by dynamical recoil, received as an effect of three-body encounters. Recoils get stronger and stronger, as the orbital separation decreases \cite{millerhamilton2002}. The BH binary will avoid ejection by dynamical recoil only if it is sufficiently massive ($\gtrsim{}50$ M$_\odot$ for a dense globular cluster, \cite{colpi2003}). If the BH binary is ejected, the loop breaks and no IMBH is formed.

Second and even more important, the merger of two BHs involves a relativistic kick. This kick may be as large as hundreds of km s$^{-1}$ \cite{lousto2009}, leading to the ejection of the BH from the parent star cluster (unless it is a nuclear star cluster). Also in this case, the loop breaks and no IMBH is formed.

Finally, even if the BH binary is not ejected, this scenario is relatively inefficient: if the seed BH is $\sim{}50$ M$_\odot$, several Gyr are required to form an IMBH with mass $\sim{}500$ M$_\odot$ \cite{millerhamilton2002}.

\begin{figure}
\center{
\includegraphics[width=12cm]{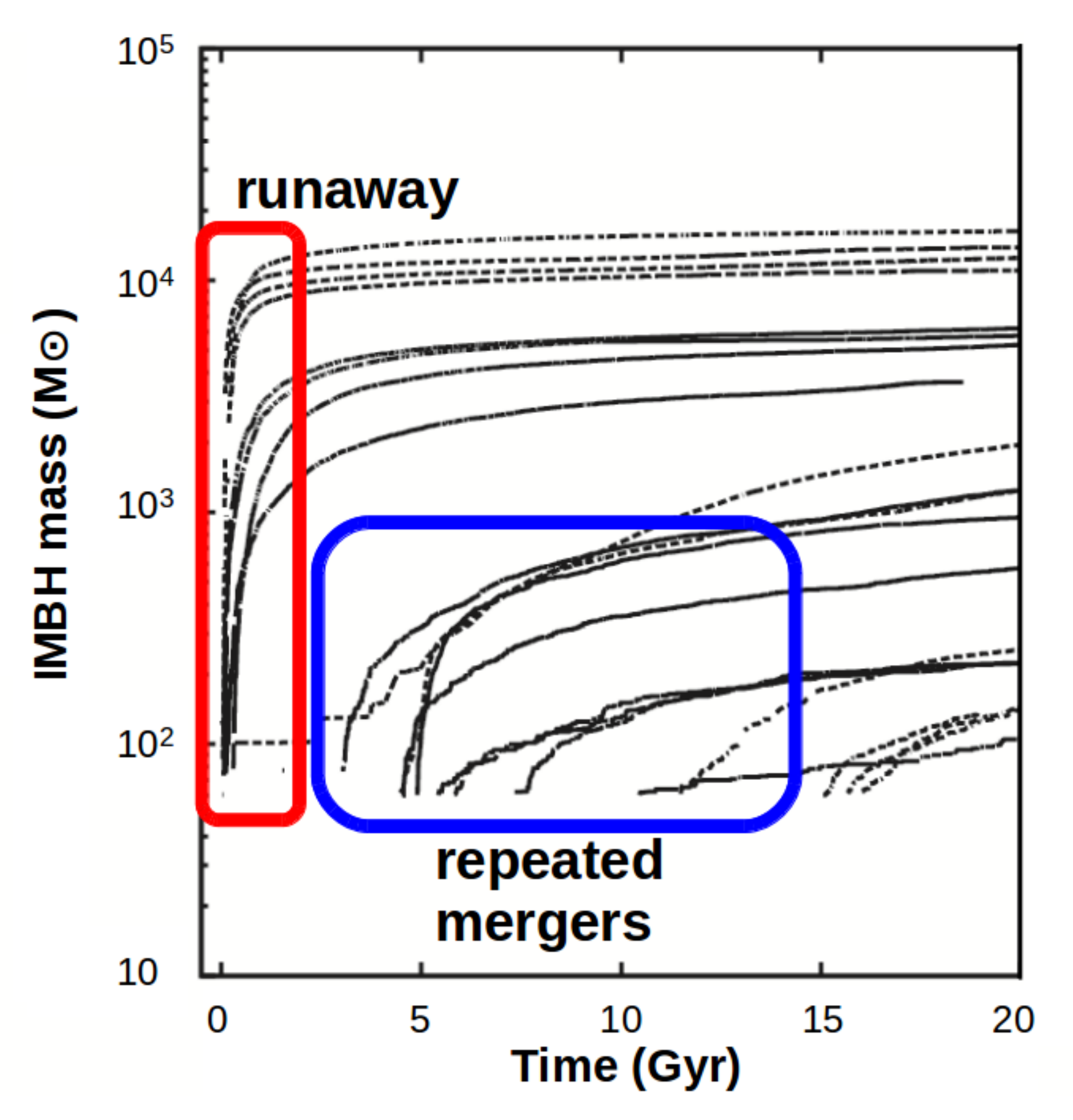}     % includes figure foo.eps
}
\caption{\label{fig:mirek}IMBH mass growth as a function of time in Monte Carlo simulations of globular clusters \cite{giersz2015}. The red (blue) box highlights IMBHs formed by runaway collisions (repeated mergers). Dashed
lines: models with reduced mass accretion on to the IMBH  and reduced star expansion after merger events. Solid lines: models with the standard prescription for IMBH mass accretion rates and post-merger expansion. See Giersz et al. (2015, \cite{giersz2015}) for further details. Adapted from Figure~6 of \cite{giersz2015}.
}
\end{figure}

Monte Carlo simulations by Giersz et al. (2015, \cite{giersz2015}) show that both the runaway collision scenario and the repeated merger scenario can be at work in star clusters. Figure~\ref{fig:mirek} clearly shows that two classes of IMBHs form in simulations: (i) runaway collision IMBHs form in the first few Myr of the life of a star cluster and grow in mass very efficiently; (ii) repeated-merger IMBHs start forming much later ($\gtrsim{}5$ Gyr) and their growth is less efficient. It is important to note that these simulations do not include prescriptions for GW recoils, which might dramatically suppress the formation of IMBHs by repeated mergers \cite{holleybockelmann2008}.
%%%%%%%%%%%%%%%%%%%%%%%%%%%%%FIGURE %%%%%%%%%%%%%%%%%%%%%%%%%%%%%%%%%%%%%%%%
%%%%%%%%%%%%%%%%%%%%%%%%%%%%%%%%%%%%%%%%%%%%%%%%%%%%%%%%%%%%%%%%%%%%%%%%%%%%

\subsection{Alternative models for IMBH formation in galactic nuclei}
Several additional models predict the formation of IMBHs in galactic nuclei. For example, \cite{millerdavies2012} propose that IMBHs can efficiently grow in galactic nuclei from runaway tidal capture of stars, provided that the velocity dispersion in the nuclear star cluster is $\gtrsim{}40$ km s$^{-1}$. Below this critical velocity, primordial binaries can support the system against core collapse, quenching the growth of the central density and leading to the ejection of the most massive BHs. Above this velocity threshold, the stellar density can grow sufficiently fast to enhance tidal captures and BH--star collisions. Tidal captures are more efficient than BH--star collisions in building up IMBHs, because the mass growth rate of the former scale as $\dot{m}_{\rm IMBH}\propto{m_{\rm IMBH}}^{4/3}$ (where $m_{\rm IMBH}$ is the IMBH seed mass), whereas the mass growth of the latter scale as $\dot{m}_{\rm IMBH}\propto{m_{\rm IMBH}}$ \cite{stone2017}. 

Furthermore, McKernan et al. (2012, \cite{mckernan2012}, see also \cite{mckernan2014}) suggest that IMBHs could efficiently grow in the accretion disc of a SMBH. Nuclear star cluster members trapped in the accretion disk are subject to two competing effects: orbital excitation due to dynamical heating by encounters with other stars and orbital damping due to gas drag. Gas damping is expected to be more effective than orbital excitation, quenching the relative velocity between nuclear cluster members and enhancing the collision rate. This favours the growth of IMBHs via both gas accretion and multiple stellar collisions. This mechanism might be considered as a gas-aided runaway collision scenario. 

%In addition to the possibility of growing IMBHs, the evolution of stellar black holes trapped in the accretion disc of a SMBH is important also because 

\subsection{Kozai-Lidov resonance}
Unlike the other dynamical processes discussed so far, Kozai-Lidov (KL) resonance \cite{kozai1962,lidov1962} can occur both in the field and in star clusters. KL resonance appears whenever we have a stable hierarchical triple system (i.e. a triple composed of an inner binary and an outer body orbiting the inner binary), in which the orbital plane of the outer body is inclined with respect to the orbital plane of the inner binary. Periodic perturbations induced by the outer body on the inner binary cause (i) the eccentricity of the inner binary and (ii) the inclination between the orbital plane of the inner binary and that of the outer body to oscillate.  

\begin{figure}
\center{
\includegraphics[width=14cm]{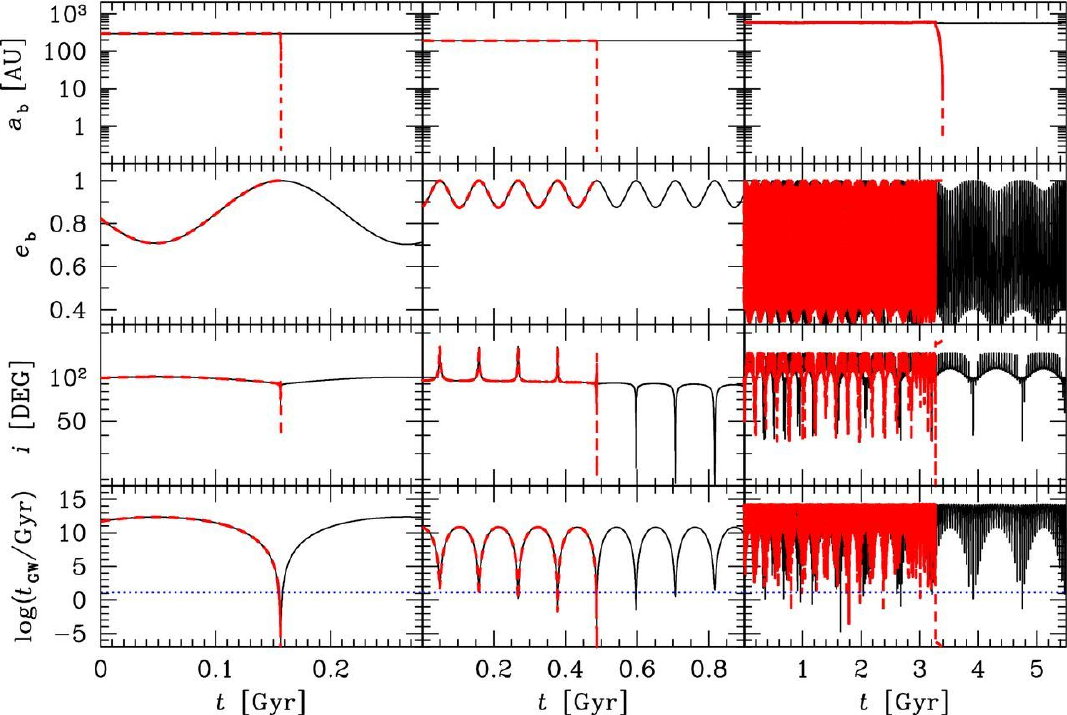}     % includes figure foo.eps
}
\caption{\label{fig:kimpson2016}Simulation of three hierarchical triples undergoing Kozai-Lidov (KL) resonance. Black solid line: simulation without post-Newtonian terms. Red dashed line: simulation with 2.5 Post-Newtonian term. From top to bottom: time evolution of semi-major axis, of the eccentricity of the inner binary, of the inclination between the orbital plane of the inner and outer binary, and of the coalescence timescale by GW emission (expressed as in \cite{peters1964}). From Figure~5 of \cite{kimpson2016}.}
\end{figure}

Figure~\ref{fig:kimpson2016} is an example of KL oscillations in three-body simulations. It is worth noting that the semi-major axis is not affected, because KL resonance does not imply an energy exchange between inner and outer binary.

%%%%%%%%%%%%%%%%%%%%%%%%%%%%%FIGURE %%%%%%%%%%%%%%%%%%%%%%%%%%%%%%%%%%%%%%%%
%%%%%%%%%%%%%%%%%%%%%%%%%%%%%%%%%%%%%%%%%%%%%%%%%%%%%%%%%%%%%%%%%%%%%%%%%%%%

KL oscillations may enhance BH binary mergers because the timescale for merger by GW emission strongly depends on the eccentricity $e$ of the binary \cite{peters1964}:
\begin{equation}
t_{\rm GW}=\frac{5}{256}\,{}\frac{c^5\,{}a^4\,{}(1-e^2)^{7/2}}{G^3\,{}m_1\,{}m_2\,{}(m_1+m_2)}.
\end{equation}

It might seem that hierarchical triples are rather exotic systems. This is not the case. In fact, $\sim{}10$ per cent of low-mass stars are in triple systems \cite{tokovinin2008,tokovinin2014,raghavan2010}. This fraction gradually increases for more massive stars \cite{duchene2013}, up to $\sim{}50$ per cent for B-type stars \cite{remageevans2011,sana2014,moe2016,toonen2017}. In star clusters, stable hierarchical triple systems may form dynamically, via four-body or multiple-body encounters.

Kimpson et al. (2016, \cite{kimpson2016}) find that KL resonance may enhance the BH merger rate by $\approx{}40$ per cent in young star clusters and open clusters. On the other hand, Antonini et al. (2017, \cite{antonini2017}) find that KL resonance in field triples can account for $\lesssim{}3$ mergers Gpc$^{-3}$ yr$^{-1}$. 

\begin{figure}
\center{
\includegraphics[width=12cm]{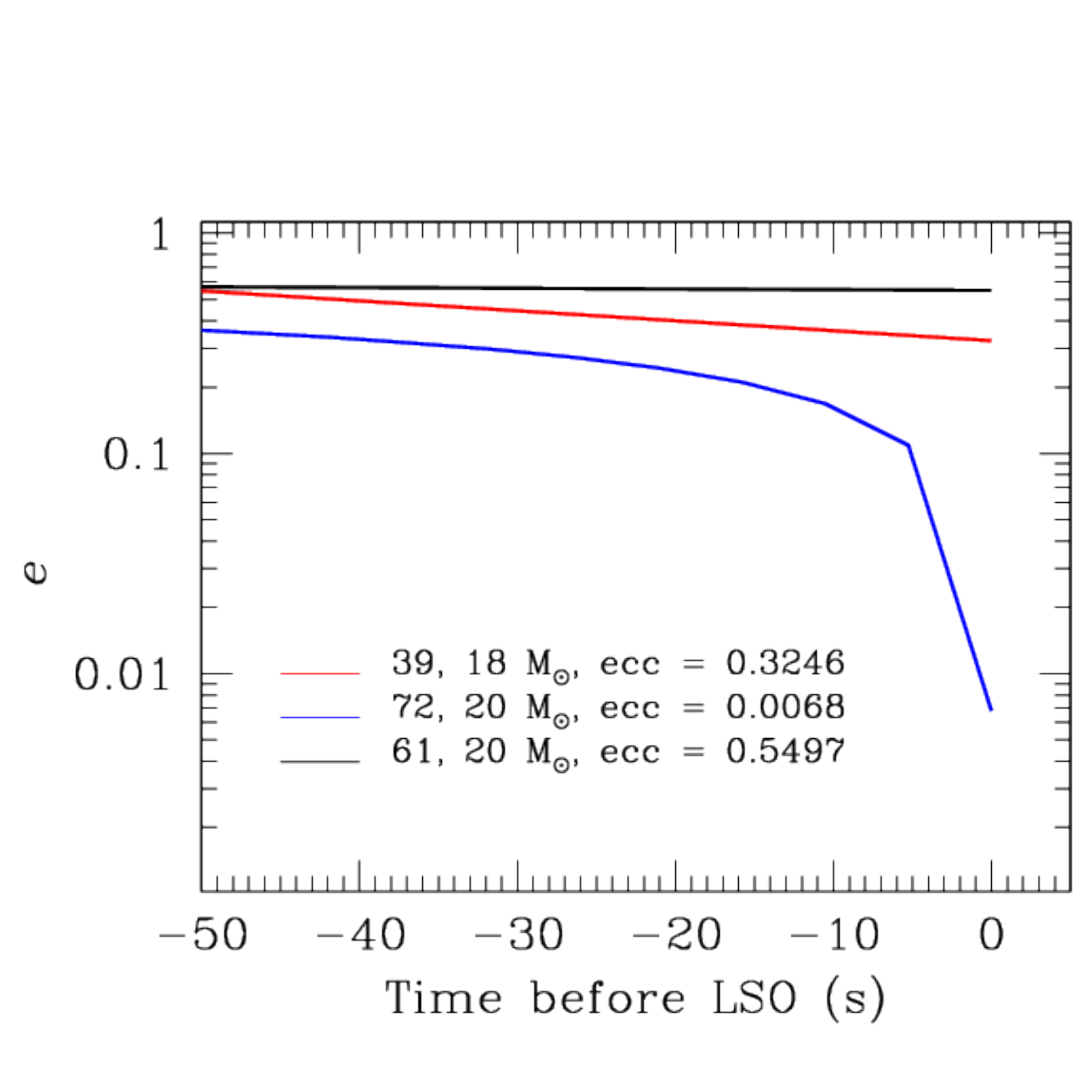}     % includes figure foo.eps
}
\caption{\label{fig:kimpson2} Evolution of the eccentricity in the last few seconds before the last stable orbit (LSO) of the three hierarchical triples already shown in Figure~\ref{fig:kimpson2016}. The LSO was estimated as $a_{\rm LSO}=6\,{}G\,{}(m_1+m_2)/c^2$. The eccentricity at the LSO is $e=0.3246$ for the system with $m_1,\,{}m_2=39,\,{}18$ M$_\odot$, $e=0.0068$ for $m_1,\,{}m_2=72,\,{}20$ M$_\odot$, and $e=0.5497$ for $m_1,\,{}m_2=61,\,{}20$ M$_\odot$.}
\end{figure}

The main signature of the merger of a KL system is the non-zero eccentricity until very few seconds before the merger. Eccentricity might be significantly non-zero even when the system enters the LIGO-Virgo frequency range, as shown in Fig.~\ref{fig:kimpson2}.

%%%%%%%%%%%%%%%%%%%%%%%%%%%%%FIGURE %%%%%%%%%%%%%%%%%%%%%%%%%%%%%%%%%%%%%%%%
%%%%%%%%%%%%%%%%%%%%%%%%%%%%%%%%%%%%%%%%%%%%%%%%%%%%%%%%%%%%%%%%%%%%%%%%%%%%

KL resonances have an intriguing application in nuclear star clusters. If the stellar BH binary is gravitationally bound to the super-massive BH (SMBH) at the centre of the galaxy, then we have a peculiar triple system where the inner binary is composed of the stellar BH binary and the outer body is the SMBH \cite{antoniniperets2012}. Also in this case, the merging BH has good chances of retaining a non-zero eccentricity till it emits GWs in the LIGO-Virgo frequency range.

\subsection{Summary of dynamics and open issues}
In this section, we have seen that dynamics is a crucial ingredient to understand BH demography. Dynamical interactions (three and few body close encounters) can favour the coalescence of BH binaries through dynamical hardening. New BH binaries can form via dynamical exchanges. Both processes suggest a boost of the BH binary merger rate in a dynamically active environment.

Moreover, exchanges favour the formation of more massive binaries, with higher initial eccentricity and with misaligned spins. Also, KL resonances favour the coalescence of more massive binaries and with higher eccentricity, even close to the last stable orbit.

On the other hand, three-body encounters might trigger the ejection of compact-object binaries from their natal environment, inducing a significant displacement between the birth place of the binary and the location of its merger.

Finally, dynamics can lead to the formation of IMBHs, with mass of few hundreds solar masses.

%%%%%%%%%%%%%%%%%%%%%%%%%%%%%FIGURE %%%%%%%%%%%%%%%%%%%%%%%%%%%%%%%%%%%%%%%%
\begin{figure}
\center{
\includegraphics[width=12cm]{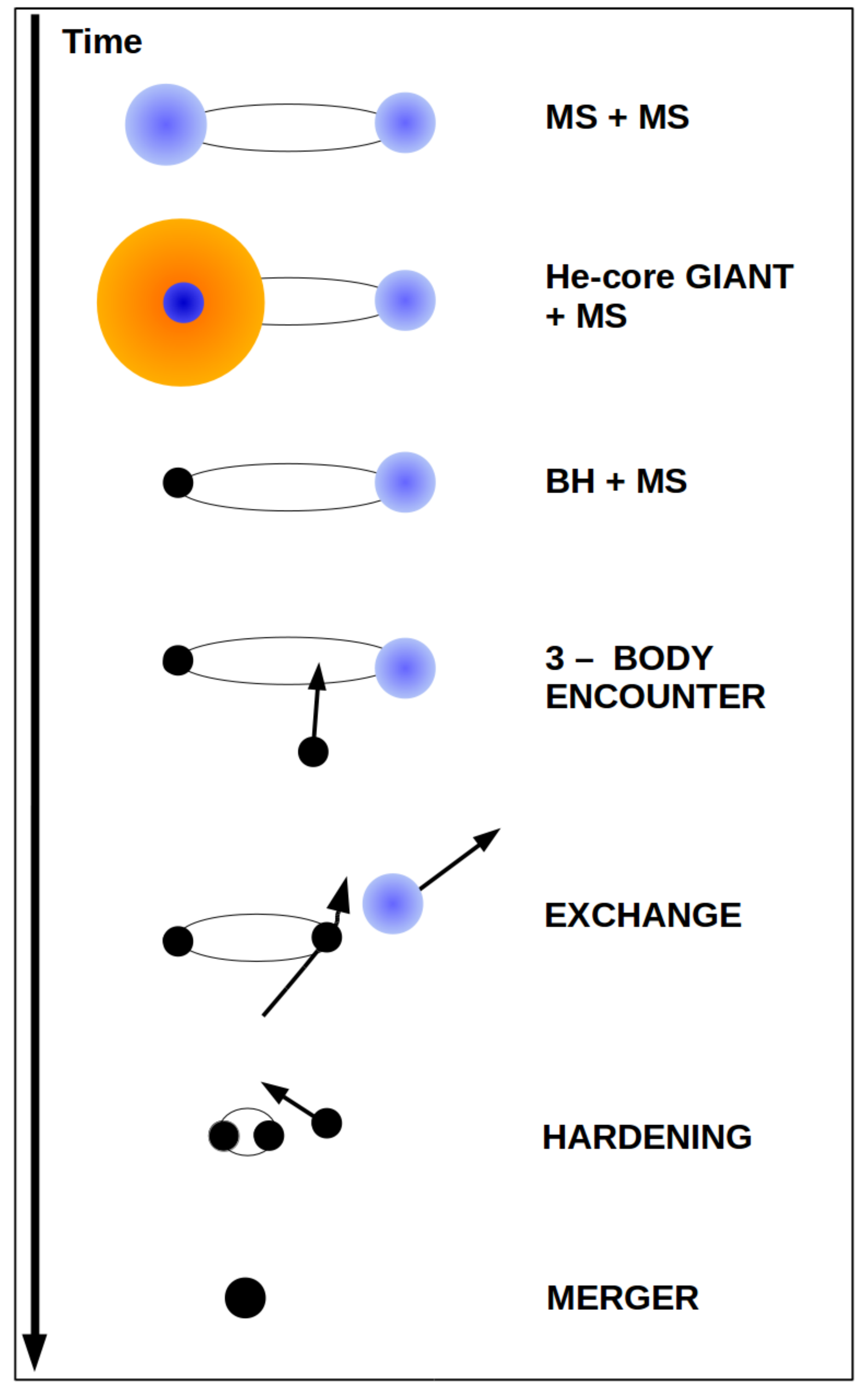}    
}
\caption{\label{fig:cartoonDYN} Schematic evolution of a merging BHB formed by dynamical exchange (see e.g. \cite{downing2010,downing2011,ziosi2014,mapelli2016,rodriguez2015,rodriguez2016,askar2017}).
}
\end{figure}
%%%%%%%%%%%%%%%%%%%%%%%%%%%%%%%%%%%%%%%%%%%%%%%%%%%%%%%%%%%%%%%%%%%%%%%%%%%%

Figure~\ref{fig:cartoonDYN} summarizes one of the possible evolutionary pathways of merging BHBs which originate from dynamics (the variety of this formation channel is too large to account for all dynamical channels mentioned above in a single cartoon). As in the isolated binary case, we start from a stellar binary. In the dynamical scenario, it is not important that this binary evolves through Roche lobe or CE (although this may happen). After the primary has turned into a BH, the binary undergoes a dynamical exchange: the secondary is replaced by a massive BH and a new BHB forms. The new binary is not ejected from the star cluster and undergoes further three-body encounters. As an effect of these three-body encounters the binary hardens enough to enter the regime in which GW emission is efficient: the BHB merges by GW decay.

We expect dynamics to be important for BH binaries also because massive stars (which are the progenitors of BHs) form preferentially in young star clusters \cite{portegieszwart2010}, which are dynamically active places. Despite this, the vast majority of studies of BH demography either do not include dynamics or focus only on old globular clusters. Globular clusters are indeed systems where dynamics is tremendously important for BH binaries, but they represent a small fraction of the stellar mass in the Universe (less than $\sim{}$ few per cent, \cite{harris2013}). 

In contrast, young star clusters are the most common birthplace of massive stars, but only few works focus on BH binaries in young star clusters \cite{ziosi2014,mapelli2016,banerjee2017a,banerjee2017b,fujii2018}. Even these studies assume overly simplified initial conditions for their simulations. For example, they do not include gas, which is a fundamental component of young star clusters. The main reason for this ``omission'' is purely numerical. Old massive globular clusters are relaxed (nearly) spherically symmetric structures with no gas. %For most of their life, they host only low mass stars and compact remnants.
They can be evolved with Monte Carlo codes (e.g. MOCCA, \cite{hypki2013}), which are considerably faster than direct N-body codes. In contrast, young star clusters are asymmetric, not yet relaxed, rich of gas and of substructures. Direct N-body codes, which scale as $N^2$, are required to model young star clusters. Adding gas to the general picture requires to combine expensive direct N-body simulations with hydrodynamical simulations. Only few authors attempted to model young star clusters with both direct N-body and hydrodynamics, in most cases simply by dividing the integration in two separate parts: first the hydrodynamics and then the direct N-body dynamics, after most gas has been converted into stars \cite{moeckel2010,pelupessy2012,fujii2015,fujii2016,parker2013,parker2015a,parker2015b,parker2017,mapelli2017}. Last but not least, young star clusters host short-lived massive stars. Thus, stellar evolution processes are very fast in young star clusters and they have a dramatic impact on the (hydro)dynamics of the system. A heroic theoretical effort is needed to properly model young star clusters and their impact on the demography of BH binaries.

\section{Black hole binaries in the cosmological context}
Population-synthesis codes model BH binaries as isolated systems, with essentially no information on the environment. Even dynamical simulations are restricted to a limited environment: the parent star cluster. On the other hand, BH binaries merging within the LIGO-Virgo instrumental horizon (all published BH mergers occurred at redshift $z\sim{}0.1-0.2$) might have formed at much higher redshift. Moreover, third-generation ground-based gravitational wave detectors (e.g. Einstein Telescope, \cite{punturo2010}) will be able to observe merging BH binaries up to redshift $z\approx{}10$. Thus, the cosmological framework in which BH binaries and their progenitors evolve cannot be neglected.

Accounting for cosmology in models of stellar-size BH binaries appears as a desperate challenge, because of the humongous dynamical range: the orbital separations of BH binaries merging within a Hubble time are of the order of tens of solar radii, while cosmic structures are several hundreds of Mpc. Several theoretical studies have faced this challenge, adopting two different procedures. 

\subsection{Analytic prescriptions}Some authors (e.g. \cite{dominik2013,dominik2015,belczynski2016,lamberts2016,dvorkin2016,dvorkin2018,giacobbo2018b}) combine the outputs of population synthesis codes with analytic prescriptions. The main ingredients are the cosmic star formation rate density and the average evolution of metallicity with redshift \cite{madau2014}. In some previous work (e.g. \cite{dominik2013,lamberts2016}) a Press-Schechter like formalism is adopted, to include the mass of the host galaxy in the general picture. Lamberts et al. (2016, \cite{lamberts2016}) even include a redshift-dependent description for the mass-metallicity relation (hereafter MZR), to account for the fact that the mass of a galaxy and its observed metallicity are deeply connected. The main advantage of this procedure is that the star formation rate and the metallicity evolution can be derived more straightforwardly from the data. The main drawback is that it is extremely difficult, if not impossible, to trace the evolution of the host galaxy of the BH binary, through its galaxy merger tree.

%%%%%%%%%%%%%%%%%%%%%%%%%%%%%FIGURE %%%%%%%%%%%%%%%%%%%%%%%%%%%%%%%%%%%%%%%%
%\begin{figure}
%\center{
%%\includegraphics[width=8cm]{dominik}     % includes figure foo.eps
%}
%\caption{\label{fig:dominik} 
%}
%\end{figure}
%%%%%%%%%%%%%%%%%%%%%%%%%%%%%%%%%%%%%%%%%%%%%%%%%%%%%%%%%%%%%%%%%%%%%%%%%%%%

\subsection{Cosmological simulations}The alternative approach feeds the outputs of population-synthesis simulations into cosmological simulations \cite{oshaughnessy2017,schneider2017,mapellietal2017,mapelli2018,mapelli2018b,cao2018}, through a Monte Carlo approach. This has the clear advantage that the properties of the host galaxies can be easily reconstructed across cosmic time. However, the ideal thing would be to have a high-resolution cosmological simulation (sufficient to resolve also small dwarf galaxies) with a box as large as the instrumental horizon of the GW detectors. This is obviously impossible. High-resolution simulations have usually a box of few comoving Mpc$^3$, while simulations with a larger box cannot resolve dwarf galaxies. Moreover, this procedure requires to use the cosmic star formation rate density and the redshift-dependent MZR which are intrinsic to the cosmological simulations. While most state-of-the-art cosmological simulations reproduce the cosmic star formation rate density reasonably well, the MZR is an elusive feature, creating more than a trouble even in the most advanced cosmological simulations.

Mapelli et al. (2017, \cite{mapellietal2017}) and Mapelli \&{} Giacobbo (2018, \cite{mapelli2018}) combine their population-synthesis simulations with the Illustris cosmological box \cite{vogelsberger2014a,vogelsberger2014b,nelson2015}. The size of the Illustris (length $=106.5$ comoving Mpc) is sufficient to satisfy the cosmological principle, but galaxies with stellar mass $\lesssim{}10^8$ M$_\odot$ are heavily under-resolved. The Illustris matches the cosmic star formation rate density and recovers a MZR. However, the simulated MZR is sensibly different from the observed relation: the curve is relatively steeper at low metallicity and there is no flattening at high mass \cite{vogelsberger2013,torrey2014}. Uncertainties connected with the MZR are estimated to affect the BH merger rate by $\sim{}20$ per cent. Keeping these {\it caveats} in mind, \cite{mapellietal2017} and \cite{mapelli2018} are able to reconstruct the BH merger rate across cosmic time. %Despite these limitations, the Illustris is one of the few large state-of-the-art cosmological simulations that are suitable for reconstructing the merger history of BH binaries (see also the EAGLE, \cite{joop2015?}).

%%%%%%%%%%%%%%%%%%%%%%%%%%%%%FIGURE %%%%%%%%%%%%%%%%%%%%%%%%%%%%%%%%%%%%%%%%
\begin{figure}
\center{
\includegraphics[width=12cm]{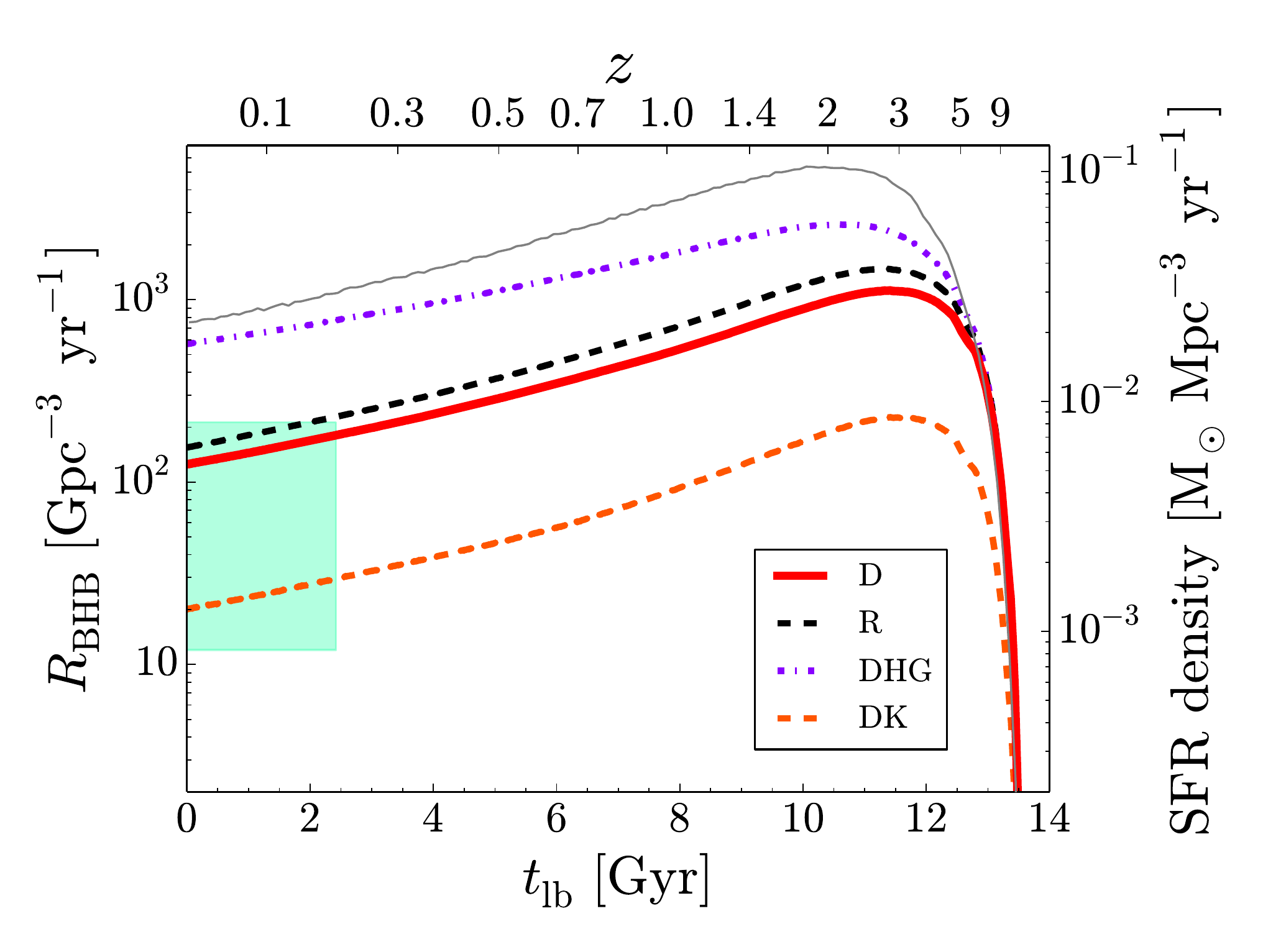}     % includes figure foo.eps
\caption{\label{fig:cosmicrate} Cosmic merger rate density of BH binaries in the comoving frame ($R_{\rm BHB}$) as a function of the look-back time ($t_{\rm lb}$, bottom $x-$axis) and of the redshift ($z$, top $x-$axis) for several population synthesis models \cite{giacobbo2018} interfaced with the Illustris simulation. Red solid line: model D (delayed SN model \cite{fryer2012}, CE parameter $\alpha{}=1$, fiducial distribution for the natal kicks of the BHs). Black dashed line: model R, same as model D but for a rapid SN model \cite{fryer2012}. Violet dot-dashed line: model DHG (same as model D but assuming that Hertzsprung gap donors can survive a CE phase). Orange dashed line: model DK (same as model D but with maximum BH natal kicks). Right-hand $y-$axis: cosmic star formation rate density in the Illustris, shown as a thin grey line. Sea-green shaded area in the left-hand part of the plot: merger rate inferred from the first three GW detections (GW150914, GW151226 and GW170104, \cite{abbott2017a}). Adapted from Figure~1 of \cite{mapellietal2017}.}}
\end{figure}
%%%%%%%%%%%%%%%%%%%%%%%%%%%%%%%%%%%%%%%%%%%%%%%%%%%%%%%%%%%%%%%%%%%%%%%%%%%%

Figure~\ref{fig:cosmicrate} shows the cosmic merger rate of BH binaries in the comoving frame obtained by \cite{mapellietal2017} for several different population-synthesis models (the assumptions for CE treatment, natal kick distribution and SN recipes are changed from model to model). In all considered models, the cosmic BH merger rate has a peak at relatively large redshift ($z\sim{}2-4$) and then decreases slowly down to current time. The trend of the cosmic merger rate is quite similar to the trend of the cosmic star formation rate density curve, modulated by both metallicity evolution and time delay\footnote{We define as time delay the time elapsed between the formation of the progenitor binary and the merger of the two BHs.}.

Mapelli et al. (2018, \cite{mapelli2018b}) also investigate the main properties of the host galaxies of merging BHs with the same Monte-Carlo approach. BHs merging in the local Universe ($z<0.024$) appear to have formed in galaxies with relatively small stellar mass ($<10^{10}$ M$_\odot$) and relatively low metallicity ($Z\leq{}0.1$ Z$_\odot$). These BHs reach coalescence either in the galaxy where they formed or in larger galaxies (with stellar mass up to $\sim{}10^{12}$ M$_\odot$). In fact, most BHs reaching coalescence in the local Universe appear to have formed in the early Universe ($\gtrsim{}9$ Gyr ago), when metal-poor galaxies were more common. A significant fraction of these high-redshift metal-poor galaxies merged within larger galaxies before the BHBs reached coalescence by GWs.

Schneider et al. (2017, \cite{schneider2017}) adopt a complementary approach to study the importance of dwarf galaxies for GW detections. They use the GAMESH pipeline to produce a high-resolution simulation of the Local Group (length = 4 Mpc comoving). This means that the considered portion of the Universe is strongly biased, but the resolution is sufficient to investigate BH binaries in small ($\gtrsim{}10^6$ M$_\odot$) dwarf galaxies. One of their main conclusions is that GW150914-like events originate mostly from small metal-poor galaxies.

Finally,  Cao et al. (2018, \cite{cao2018}) investigate the host galaxies of BHBs by applying a semi-analytic model to the Millennium-II N-body simulation \cite{boylan2009}. The Millennium-II N-body simulation is a large-box (length = 137 comoving Mpc) dark-matter only simulation. The physics of baryons is implemented  through a semi-analytic model. Using a dark-matter only simulation coupled with a semi-analytic approach offers the possibility of improving the resolution significantly, but baryons are added only in post-processing. With this approach, \cite{cao2018} find that BHBs merging at redshift $z\lesssim{}0.3$ are located mostly in massive galaxies (stellar mass $\gtrsim{}2\times{}10^{10}$ M$_\odot$). 

These studies show that the combination of population-synthesis tools with cosmological simulations is a crucial approach to understand the cosmic evolution of the merger rate and the properties of the host galaxies of BH mergers. % comparing the results of large-box cosmological simulations with small-box high-resolution simulations is surely crucial to understand how BH binaries populate the galaxies across cosmic time.
%\begin{figure}
%\includegraphics{foo}     % includes figure foo.eps
%\caption{Description of the figure.}
%\end{figure}

%Tables~\ref{tab:pricesI}

%\begin{table}
%  \caption{Prices of important items.}
%  \label{tab:pricesI}
%  \begin{tabular}{rcl}
%    \hline
%      Ice-cream      & 1500  & lire    \\
%      More ice-cream & 15000 & lire    \\
%      Crocodile      & 1500  & dollars \\
%    \hline
%      Phone call     & .25   & dollars \\
%      X-Men          & 1.25  & dollars \\
%      Dollar         & 1     & dollars \\
%    \hline
%  \end{tabular}
%\end{table}

\section{Summary and outlook}
%{\bf to be written}
We reviewed our current understanding of the astrophysics of stellar-mass BHs.
 The era of gravitational wave astrophysics has just begun and has already produced two formidable results: BH binaries exist and can host BHs with mass $>30$ M$_\odot$ \cite{abbott2016a,abbott2016d}.

 According to nowadays stellar evolution and supernova theories, such massive BHs can form only from massive relatively metal-poor stars. At low-metallicity, stellar winds are quenched and stars end their life with a larger mass than their metal-rich analogues. If its final mass and its final core mass are sufficiently large, a star can directly collapse to a BH with mass $\gtrsim{}30$ M$_\odot$ \cite{mapelli2009,belczynski2010}. An alternative scenario predicts that $\sim{}30-40$ M$_\odot$ BHs are the result of gravitational instabilities in the very early Universe (primordial BHs, e.g. \cite{carr2016}). 

The formation channels of merging BH binaries are still an open question. All proposed scenarios have several drawbacks and uncertainties. While mass transfer and common envelope are a major issue in the isolated binary evolution scenario, even the dynamical evolution is still effected by major issues (e.g. the small statistics about BHs in young star clusters, and the major simplifications adopted in dynamical simulations).

Finally, a global picture is missing, which combines stellar and binary evolution with dynamics and cosmology, aimed at reconstructing the BH merger history across cosmic time. This is crucial for the astrophysical interpretation of LIGO-Virgo data and for meeting the challenge of third-generation ground-based GW detectors. 

\acknowledgments
I thank the organizers and the participants of the International School of Physics ``Enrico Fermi'' (Course 200 - Gravitational Waves and Cosmology) for the enlightening discussions and for giving me the opportunity to think about these lectures. Numerical calculations have been performed through a CINECA-INFN agreement and through a CINECA-INAF agreement (Accordo Quadro INAF-CINECA 2017), providing access to resources on GALILEO and MARCONI at CINECA.
 MM  acknowledges financial support from the MERAC Foundation through grant `The physics of gas and protoplanetary discs in the Galactic centre', from INAF through PRIN-SKA `Opening a new era in pulsars and compact objects science with MeerKat', from MIUR through Progetto Premiale 'FIGARO' (Fostering Italian Leadership in the Field of Gravitational Wave Astrophysics) and 'MITiC' (MIning The Cosmos  Big Data and Innovative Italian Technology for Frontier Astrophysics and Cosmology), and from the Austrian National Science Foundation through FWF stand-alone grant P31154-N27 `Unraveling merging neutron stars and black hole - neutron star binaries with population-synthesis simulations'. This work benefited from support by the International Space Science Institute (ISSI), Bern, Switzerland,  through its International Team programme ref. no. 393
 {\it The Evolution of Rich Stellar Populations \& BH Binaries} (2017-18).
%MM  acknowledges financial support from the Italian Ministry of Education, University and Research (MIUR) through grant FIRB 2012 RBFR12PM1F, from INAF through grant PRIN-2014-14, and from the MERAC Foundation.

\end{document}